\documentclass[aps,prb,twocolumn,showpacs,floatfix]{revtex4-2}

\usepackage{graphicx,color}
\usepackage{amsfonts}
\usepackage[figuresright]{rotating}
\usepackage{amssymb}
\usepackage{amsmath}
\usepackage{pifont}
\usepackage{psfrag}
\usepackage{subfigure}
\usepackage{multirow}
\usepackage{tabularx}
\usepackage{textcomp}
\usepackage{units}
\usepackage{hyperref}
\usepackage{comment}
\usepackage{longtable}
\usepackage{tikz}
\usepackage{circledsteps}

\hypersetup{
 pdfnewwindow=true, colorlinks=true,
 linkcolor=blue, anchorcolor=blue,
 citecolor=blue, filecolor=blue,
 menucolor=blue, urlcolor=blue}

\def\nn{\nonumber}

\def\beq{\begin{eqnarray}}
\def\eeq{\end{eqnarray}}

\renewcommand{\v}[1]{\ensuremath{\mathbf{#1}}} 

\let\baraccent=\= 
\renewcommand{\=}[1]{\stackrel{#1}{=}} 
	
\makeatletter

\begin{document}

\title{Average density of Bloch electrons in a homogeneous magnetic field: A second-order response}

\author{Benjamin\ \surname{M. Fregoso}}
\affiliation{Department of Physics, Kent State University, Kent, Ohio 44242, USA}

\begin{abstract}
We compute the average density of a three-dimensional multiband crystal of arbitrary symmetry, metal or insulator, to first and second order in a weak homogeneous magnetic field. To linear order and for insulators, the density follows the well-known Streda formula, but for metals there is an extra contribution from the orbital magnetic moments at the Fermi surface. To second order the average density depends on several microscopic processes. Among these, the quantum metric tensor plays an important role by generating a pseudo-magnetic moment resulting from the rotation of the Bloch wave functions in the complex projective plane. We also discuss the  implications of our results for the volume and pressure. The method we develop is explicitly gauge invariant, considers intraband and interband processes on equal footing, accommodates relaxation processes, and can be readily extended to other observables.
\end{abstract}

\maketitle

\section{Introduction and main results}
Electronic systems in magnetic fields give rise to many novel phenomena~\cite{Ashcroft1976,Ziman2001,Nagaosa2010,Widom1982,Streda1982,MacDonald1984,Xu2025,Huang2025,Xiao2005,Vanderbilt2018,Son2012,PeraltaGavensky2023,Fregoso2024,PeraltaGavensky2025,Hetenyi2025,Hetenyi2024,Atencia2024,Pereira2009,Levy2010,Zhang2018,Roy2014,Shi2007}. Strong-field responses have long inspired new ideas, particularly in connection with quantum Hall effects, but the weak-field regime has received comparatively less attention. In this work, we study how the average electronic density is modified by the presence of a weak static homogeneous magnetic field in a clean three-dimensional (3D) crystal of arbitrary symmetry, whether metal or insulator. We focus on the regime in which the field is strong enough to trigger interesting nonlinear responses~\cite{Fregoso2019,Fregoso2022,Mendoza2024,Oike2024,Ezawa2025,Kirikoshi,Zhou2025,Sarkar2025,Jiang2025} but weak enough that Bloch wave functions remain a valid description of the electronic states. 

Generally, the average particle density can be expanded in powers of the magnetic field as 
\begin{align}
n = n^{(0)} + \varsigma^{(1)a} B^{a} + \varsigma^{(2)ab} B^{a}B^{b} + \cdots. 
\label{eq:n_powers_B}
\end{align}
The first term on the right-hand side represents the density in the absence of magnetic field, the second term is the linear response, and the third is the nonlinear response. Since the density is even under time reversal while the magnetic field is odd, the linear response vanishes if time reversal symmetry (TRS) is present. 

If TRS is broken, the linear response is dominant, and we expect the (pseudo)tensor $\varsigma^{(1)}$ to follow the well-known Streda formula~\cite{Widom1982,Streda1982,MacDonald1984}  
\begin{align}
\left(\frac{\partial n}{\partial \v{B}}\right)_{\mu}= -\frac{1}{e} \pmb{\sigma}_{H},
\label{eq:streda}
\end{align}
which relates the average density to the anomalous Hall conductivity $\pmb{\sigma}_{H}$ when the chemical potential lies within an energy gap. This relation was originally derived in the context of strong magnetic fields but later found to apply across a wide range of field strengths, even in the presence of electron interactions. It predicts that the average density changes in discrete steps when the anomalous Hall conductivity is quantized with nonzero Chern number. Indeed, it has been used as a diagnostic tool to measure the Chern numbers~\cite{Xu2025,Huang2025} and has applications within the semiclassical theory of electron dynamics~\cite{Xiao2005}, orbital magnetization~\cite{Vanderbilt2018}, Fermi liquid theory~\cite{Son2012}, Chern insulators~\cite{PeraltaGavensky2023}, and out of equilibrium~\cite{PeraltaGavensky2025} and localization physics~\cite{Hetenyi2025,Hetenyi2024}, among others. 

First, we compute the response tensor $\varsigma^{(1)}$ for a system of Bloch electrons in Sec.~\ref{sec:n_first}. We find that, indeed, the Streda formula holds for Bloch electrons when the chemical potential lies within an energy gap, but there is a correction when the chemical potential crosses an energy band~\cite{Fregoso2024}, 
\begin{align}
\frac{\partial n}{\partial \v{B}}= -\frac{1}{e} \pmb{\sigma}_{H} +  \frac{1}{V}\sum_{n\v{k}} f'_{n} \v{m}_{n}.
\label{eq:streda_fregoso}
\end{align}
The notation used can be found in Appendix \ref{app:identities}. The second term arises from the interaction of the magnetic field with the orbital magnetic moments $\v{m}_{n}$ (\ref{eqn:orb_moment}) at the Fermi surface (FS). $f'_{n}$ is the first derivative of the Fermi function (\ref{eq:Fermi_func}), which is nonzero only near the Fermi level. Equation (\ref{eq:streda_fregoso}) implies that particles can be transferred to (and from) the FS and the edge states (assuming the expression can be applied in the Hall regime). 

When TRS is present, the second-order response is the leading contribution. We derive a general expression for $\varsigma^{(2)}$ in terms of Bloch wave functions in Sec.~\ref{sec:n_2nd}. It is not as compact as (\ref{eq:streda_fregoso}) since it includes FS, non-FS, one-band, two-band, and three-band terms and must generally be analyzed on a case-by-case basis. There is a contribution that deserves special attention (Sec.~\ref{sec:geometric}). We find that the Fubini-Study quantum metric tensor $g^{ab}_{n}$ which characterizes the distance between Bloch states~\cite{Provost1980,Page1987,Anandan1990,Kobayashi1969,Pati1991} induces quadratic magnetic moments such as 
\begin{align}
\mathfrak{m}_{n}^{(2)zz} \sim e^2 g_{n}^{xx} v_{n}^y v_{n}^y,  
\end{align}
where $v^a_n$ is the band velocity in direction $a$. The origin of this magnetic moment is purely geometric, meaning it does not arise from spin or orbital angular momentum but, rather, from angular momentum due to rotations of Bloch wave functions in the complex projective plane (Fig.~\ref{fig:qmt}). The quantum \textit{geometric} magnetic moment appears only in the nonlinear response and is distinct from known mechanisms that generate magnetic field-like effects without an actual magnetic field, such as those found in strained graphene\cite{Pereira2009,Levy2010,Zhang2018} or in the flat-band fractional quantum Hall effect~\cite{Roy2014}. 

In Sec.~\ref{sec:magnetovolume}, we argue that a change in density implies a change in volume or pressure (however small) since these thermodynamic qualities can also be expanded in powers of the magnetic field 
\begin{align}
V &= V^{(0)} + v^{(1)a} B^{a} + v^{(2)ab} B^{a}B^{b} + \cdots,  \\
P &= P^{(0)} + p^{(1)a} B^{a} + p^{(2)ab} B^{a}B^{b} + \cdots,  
\label{eq:v_p_powers_B}
\end{align}
and, when the chemical potential lies within an energy gap, $v^{(1)a}$ and $p^{(1)a}$ are also proportional to the anomalous Hall conductivity.  

In Sec. \ref{sec:11_21_terms} we show that in a simple two-band model the intraband contribution is largely due to geometric magnetic moments
\begin{align}
\varsigma^{(2)}_{intra} &\sim \frac{1}{V} \sum_{n\v{k}} f''_{n}\mathfrak{m}_{n}^{(2)zz},  
\end{align}
and results in a fractional density change in the order, $n^{(2)}/n^{(0)}\sim 10^{-4}\%$. More accurate values require large-scale numerical simulations that incorporate detailed crystal structure information. The analytical expressions obtained in this work are well suited for that purpose.

\section{Particle density}
\label{sec:particle density}
We consider non-interacting electrons on a 3D lattice whose single-particle Hamiltonian is 
\begin{align}
h= \frac{1}{2m}(\v{p}-e \v{A})^2 + U,
\label{eq:sp_ham}
\end{align}
where $e<0$ is the electron charge and $U$ is the ionic periodic potential. Without loss of generality, the magnetic field is taken along  the $z$ axis and represented by the vector potential $\v{A} = \hat{\v{y}} (B^z/q)\sin(\v{g}\cdot\v{r})$. The magnetic field is spatially modulated by a wave vector $\v{g}=q\hat{\v{x}}$. We introduce the parameter $q$ to regularize the singular nature of the position operator in the Bloch basis~\cite{Shi2007}. Since we are interested in the response to a homogenous magnetic field, we take the $q\to 0$ limit at the end of the calculation. In this limit, the magnetic field is homogeneous:  
\begin{align}
\v{B}=\hat{\v{z}} B^z.
\end{align}
In addition, at every step of our calculation our expressions are invariant under the phase transformation of Bloch states, $|n\v{k}\rangle \to e^{i\alpha(\v{k})}|n\v{k}\rangle$  (we also call it gauge invariance).

The particle density operator 
\begin{align}
\hat{n} = \psi^{\dagger}\psi
\label{eq:density_op_def}
\end{align}
can be expanded in Bloch operators $a_{n\v{k}}$ as $\psi=\sum_{n\v{k}} \varphi_{n\v{k}}a_{n\v{k}}$, where $\varphi_{n\v{k}}$ is a Bloch wave function with band $n$ and crystal momentum $\v{k}$. Its expectation value is then
\begin{align}
n &= \sum_{n\v{k}m\v{k}'} \varphi^{*}_{n\v{k}} \varphi_{m\v{k}'} \rho_{m\v{k}'n\v{k}}, 
\label{eqn:avgn} 
\end{align}
where $\rho_{m\v{k}'n\v{k}} \equiv \langle a_{n\v{k}}^{\dagger} a_{m\v{k}'}  \rangle$ is the density matrix in the Bloch basis (see Appendix \ref{app:densitymat}). Because the magnetic field varies spatially, it is convenient to expand the density in its spatial Fourier modes as 
\begin{align}
n &=n_0 +  n_{\v{g}} \cos(\v{g}\cdot\v{r}) + n_{2\v{g}} \cos(2\v{g}\cdot\v{r}) + \cdots,
\label{eqn:n_f_exp}
\end{align}
where 
\begin{align}
n_{\v{G}} &= \frac{2}{V}\int d\v{r}~ n \cos{( \v{G}\cdot\v{r})}~~~~~ (\v{G}\neq 0). 
\label{eqn:F_modes_cos}
\end{align}
The $\cdots$ in (\ref{eqn:n_f_exp}) represents higher-order Fourier modes. The expansion also contains $sine$ modes, but as shown in Appendix~\ref{app:sineterms}, they vanish. Here, $n_0$ is the density in the absence of a magnetic field, $n_{\v{g}}$ corresponds to the linear correction, and $n_{2\v{g}}$ corresponds to the quadratic correction. Substituting (\ref{eqn:avgn}) into (\ref{eqn:F_modes_cos}), we obtain 
\begin{align}
n_{\v{G}} &= \frac{2}{V} \textrm{Re} \sum_{nm\v{k}} \langle u_{n\v{k}} | u_{m\v{k}-\v{G}}\rangle \rho_{m\v{k}-\v{G}n\v{k}},
\label{eq:n_2}
\end{align}  
which is valid to all orders in $B^z$ and for arbitrary $\v{G}$. The average density is then
\begin{align}
n-n_0= \lim\limits_{q\to 0} \sum_{\v{G}} n_{\v{G}}, 
\end{align}
where the sum runs over $\v{G}=\v{g},2\v{g}$.

\section{Zero order}
\label{sec:n_zero}
To zero order, only $n_0$ contributes:
\begin{align}
n^{(0)} \equiv n_0=  \frac{1}{V}\sum_{n\v{k}} f_{n\v{k}},
\end{align}
where we used  $\rho^{(0)}_{n\v{k}n\v{k}}=f_{n\v{k}}$.

\section{First order}
\label{sec:n_first}
To first order, only $n_{\v{g}}$ contributes: 
\begin{align}
n^{(1)} &= \lim\limits_{q\to 0} n^{(1)}_{\v{g}}.
\label{eq:n0}
\end{align}
In Appendix \ref{app:densitymat}, we calculate $\rho^{(1)}$ in the \textit{static} and \textit{clean} limits ($\tau \to \infty$). Substituting this into (\ref{eq:n_2}) gives
\begin{align}
n^{(1)}_{\v{g}} &=\varsigma_{\v{g}}^{(1)a} B^{a}, 
\end{align}
where the $z$ component of the response tensor is 
\begin{align}
 \varsigma_{\v{g}}^{(1)z} &\hspace{-3pt}=\hspace{-3pt}  \frac{2}{V}\left(\frac{e}{2q}\right) \textrm{Im} \sum_{nl\v{k}}\langle u_{n\v{k}} | P_{l\v{k}-\v{g}} \hat{v}^{y}_{\v{k}} |  u_{n\v{k}}\rangle \frac{f_{l\v{k}-\v{g}} \hspace{-2pt}-\hspace{-2pt} f_{n\v{k}}}{\varepsilon_{l\v{k}-\v{g}} \hspace{-2pt}-\hspace{-2pt} \varepsilon_{n\v{k}}}.
\label{eq:vars1}
\end{align}
All quantities are defined in Appendix \ref{app:identities}: $f_{n\v{k}}$ is the Fermi function (at zero temperature) of electrons at energy $\varepsilon_{n\v{k}}$, and $P_{n\v{k}}= |u_{n\v{k}}\rangle \langle u_{n\v{k}}|$ is the projector onto band $n$ at point $\v{k}$. In deriving (\ref{eq:vars1}), we used $\hat{v}^{y}_{\v{k}-\v{g}}=\hat{v}^{y}_{\v{k}}$ because $\v{g}=\hat{\v{x}}q$ is perpendicular to $\hat{v}^{y}_{\v{k}}$; this also follows from (\ref{eq:partialvk}). Note that (\ref{eq:vars1}) is explicitly gauge invariant- the sum is the trace over a gauge-invariant product of operators. 

The operator $P_{l\v{k}-\v{g}} \hat{v}^{y}_{\v{k}}$ is interesting because its $q\to 0$ limit, $\sim(\partial_{x} P_{l}) \hat{v}^{y}$ ($\v{k}$ omitted), could be interpreted as a quantum momentum tensor (QMT) \textit{operator}. To see this, we compute its average over a  Bloch state and note that it yields all possible momentum components,
\begin{align}
 \langle u_{n} | (\partial_a P_n) \hat{v}^{b}_{\v{k}}| u_{n} \rangle &=  \langle \partial_a u_n | \frac{\varepsilon_n - h_{0\v{k}}}{\hbar}| \partial_b u_n \rangle \label{eq:qmt_v0}\\
&=\frac{1}{2} [ (\partial_a v^{b}_n) \hspace{-2pt}-\hspace{-2pt} \frac{\hbar}{m}\delta_{a,b} ] \hspace{-2pt}- \hspace{-2pt}\frac{i }{e} \epsilon_{abc} m^c_{n}.
\label{eq:qmt}
\end{align} 
To derive (\ref{eq:qmt_v0}) we used (\ref{eq:projector}) and (\ref{eqn:vk_on_unk}). To derive (\ref{eq:qmt}) we used (\ref{eqn:vk_on_unk}), (\ref{eq:inv_mass_ten}), and (\ref{eqn:orb_moment}). The first term on the right-hand side is real, symmetric in the $a,b$ indices, and proportional to the momentum. The second term is purely imaginary, anti-symmetric in the $a,b$ indices, and proportional to the orbital angular momentum; here $\v{m}_n$ denotes the standard orbital angular moment~\cite{Atencia2024} of electrons in band $n$ (\ref{eqn:orb_moment}). Clearly, only the antisymmetric part interacts with the external magnetic field, and therefore, $\v{m}_n$ appears in the final expression for $\pmb{\varsigma}^{(1)}$. The QMT is similar to the quantum geometric tensor (\ref{eq:qgeoten}), where instead of $(\varepsilon_{n\v{k}} - h_{0\v{k}})/\hbar$ we have $1-P_{n\v{k}}$:
\begin{align}
Q^{ab}_n &\equiv \langle \partial_a u_n | 1- P_n | \partial_b u_n \rangle = g^{ab}_n -\frac{i}{2} \Omega^{ab}_{n}.
\label{eq:qgt}
\end{align} 
Its real part corresponds to the Fubini--Study quantum metric tensor (\ref{eq:qgeoten2}), and its imaginary part corresponds to the Berry potential (\ref{eqn:berry_potential}). 

Next, we separate $m \neq n$ from $m=n$ processes in (\ref{eq:vars1}) and take the $q\to 0$ limit. After some algebra (see Appendix \ref{sec:eqn_n1_berry_c}) we obtain
\begin{align}
\pmb{\varsigma}^{(1)} &=-\frac{e}{\hbar V}\sum_{n\v{k}} f_{n} \pmb{\Omega}_{n}\hspace{-2pt} \hspace{-1pt}+\hspace{-1pt} \frac{1}{V}\sum_{n\v{k}} f'_{n} \v{m}_{n}.
\label{eqn:n1_berry_c}
\end{align}
The first term on the right-hand side can be recognized as being proportional to the anomalous Hall conductivity (\ref{eqn:Hall_conductivity}), and hence, we obtain (\ref{eq:streda_fregoso}). This expression agrees with the Streda formula (\ref{eq:streda}) for insulators, but there is a contribution proportional to the orbital magnetic moments located at the FS since $f'_n \equiv \partial f_{n}/\partial \varepsilon_{n}$ is localized near a small neighborhood around the Fermi level~\cite{Fregoso2024,PeraltaGavensky2025}.

\section{Second order}
\label{sec:n_2nd}
To second order only $n_{2\v{g}}$ contributes:
\begin{align}
n^{(2)} &= \lim\limits_{q\to 0} n^{(2)}_{2\v{g}},
\label{eq:n2}
\end{align}
where $n^{(2)}_{2\v{g}}$ is given by (\ref{eq:n_2}). Substituting $\rho^{(2)}$ from Appendix \ref{app:densitymat}, we obtain
\begin{align}
n^{(2)}_{2\v{g}} &=\varsigma_{2\v{g}}^{(2)ab} B^{a}B^{b}, 
\end{align}
where the $zz$ component of the response tensor is 
\begin{align}
\varsigma_{2\v{g}}^{(2)zz} &= -\frac{2}{V}\hspace{-2pt}\left(\frac{e}{2iq}\right)^2\hspace{-4pt}  \textrm{Re}\hspace{-3pt}\sum_{mnl\v{k}} \frac{1}{\varepsilon_{m\v{k}\hspace{-1pt}-\hspace{-1pt}2\v{g}} \hspace{-2pt}-\hspace{-2pt}\varepsilon_{n\v{k}}} \nn \\
&~~~~~~~~~~~~~\times 
\bigg[\frac{f_{l\v{k}-\v{g}}\hspace{-2pt}-\hspace{-2pt}f_{n\v{k}}}{\varepsilon_{l\v{k}-\v{g}} \hspace{-2pt}-\hspace{-2pt}\varepsilon_{n\v{k}}} - \frac{f_{m\v{k}\hspace{-1pt}-\hspace{-1pt}2\v{g}}\hspace{-2pt}-\hspace{-2pt}f_{l\v{k}-\v{g}}}{\varepsilon_{m\v{k}\hspace{-1pt}-\hspace{-1pt}2\v{g}} \hspace{-2pt}-\hspace{-2pt}\varepsilon_{l\v{k}-\v{g}}}  \bigg] \nn \\ 
& ~~~~~~~~~~~~~\times \langle  u_{n\v{k}}|P_{m\v{k}-2\v{g}} \hat{v}^{y}_{\v{k}-\v{g}} P_{l\v{k}-\v{g}} \hat{v}^{y}_{\v{k}}|u_{n\v{k}}\rangle
\nn \\
&+\hspace{-2pt}\frac{1}{mV}\hspace{-2pt}\left(\frac{e}{2iq}\right)^2\hspace{-4pt} \textrm{Re}\hspace{-3pt}\sum_{mn\v{k}}  \frac{f_{m\v{k}\hspace{-1pt}-\hspace{-1pt}2\v{g}}\hspace{-2pt}-\hspace{-2pt}f_{n\v{k}}}{\varepsilon_{m\v{k}\hspace{-1pt}-\hspace{-1pt}2\v{g}} \hspace{-2pt}-\hspace{-2pt}\varepsilon_{n\v{k}}}\langle  u_{n\v{k}}|P_{m\v{k}\hspace{-1pt}-\hspace{-1pt}2\v{g}}| u_{n\v{k}}\rangle.  
\label{eq:n2g}
\end{align}  
As in the linear response case, expression (\ref{eq:n2g}) is explicitly gauge invariant. The first term represents processes mediated by two first-order perturbations, $(P \hat{v}^y)(P \hat{v}^y)$. After taking the $q\to 0$ limit, terms involving second derivatives appear, such as ($\partial^2_x P) \hat{v}^y$, second derivatives of Fermi functions, and many mixed products with one derivative from either pair of operators and one from a Fermi function. In general, these terms are harder to interpret and demonstrate the richness of the second-order quantum processes involved.

\begin{figure}[t]
\subfigure{\includegraphics[width=.47\textwidth]{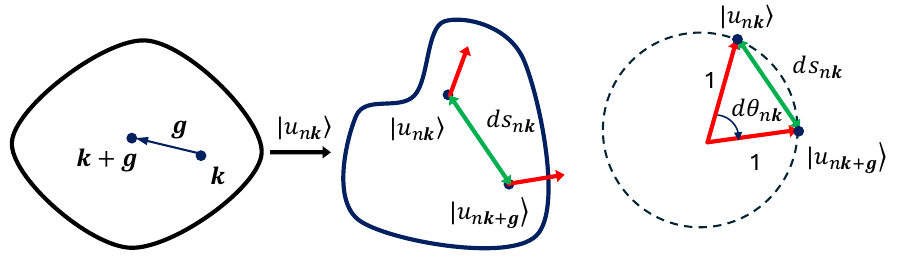}}
\caption{Schematic representation of the mapping from parameter space $\v{k}$ to the complex projective plane of Bloch wave functions. A small shift by $\v{g}$ rotates the vector $|u_{n\v{k}} \rangle$ by an angle $d\theta_{n\v{k}}$ [see Eq.~(\ref{eq:ds2})].}
\label{fig:qmt}
\end{figure}

The second term in (\ref{eq:n2g}) represents processes mediated by the quadratic (diamagnetic) perturbation. The operator $P_{n\v{k}-2\v{g}}$, which, after the $q \to 0$ limit, becomes $\sim \partial_x^2 P_{n}$, could be interpreted as the metric tensor \textit{operator} because its expectation value over a Bloch state is proportional to the quantum metric. Indeed, using (\ref{eqn:id1}) and its derivative, we obtain $\langle u_{n\v{k}} | P_{n\v{k}-2\v{g}} | u_{n\v{k}} \rangle = 1 - g_{n\v{k}}^{xx}(2q)^2 + \cdots$, and for a small momentum shift $\v{g}$ in an arbitrary direction,  
\begin{align}
 \langle u_{n\v{k}} | P_{n\v{k}-2\v{g}} | u_{n\v{k}} \rangle = 1 - g_{n\v{k}}^{ab}(2g^a)(2g^b) + \cdots,
\end{align}
where 
\begin{align}
\langle u_{n}|\partial_{ab} P_{n}|u_{n}\rangle = -2 g^{ab}_{n}.
\label{eq:metric_gab}
\end{align}
The Fubini--Study metric tensor characterizes the distance between states $|u_{n\v{k}}\rangle$ and $|u_{n\v{k}+\v{g}}\rangle$. We can understand this intuitively as follows. Since these vectors are normalized, a small shift by $\v{g}$ \textit{rotates} $|u_{n\v{k}}\rangle$ by a small angle in the complex projective plane (see Fig. \ref{fig:qmt}). The distance between these states is given by the law of cosines: $ds_{n\v{k}}^2 = 2-2\cos{(d\theta_{n\v{k}})} \approx d\theta_{n\v{k}}^2$. Defining the angle by taking the norm of the usual inner product $\langle u_{n\v{k}}|u_{n\v{k}+\v{g}} \rangle = \lVert | u_{n\v{k}}\rangle \rVert^2   \lVert |u_{n\v{k}+\v{g}} \rangle \rVert^2 \cos{(d\theta_{n\v{k}})}$ (the angle must be real), we obtain 
\begin{align}
ds_{n\v{k}}^2 = d\theta^2_{n\v{k}} = 1 - \lVert \langle u_{n\v{k}}|u_{n\v{k}+\v{g}} \rangle\rVert^2 = g_{n\v{k}}^{ab} g^a g^b. 
\label{eq:ds2}
\end{align} 
Thus, the magnetic field's only effect is to rotate the Bloch wave functions since (as expected from semiclassical dynamics) the band energies remain unchanged up to second order, $\varepsilon_{n\v{k}}^{(1)} =\varepsilon_{n\v{k}}^{(2)}=0$. The introduction of $q$ allows us to replace the position operator matrix elements (which is singular in the Bloch basis) for k-space derivatives (in the tangent bundle to the BZ) while time keeping both intra- and interband processes on equal footing. 

Another way to interpret (\ref{eq:metric_gab}) is to note that the diamagnetic part of the Hamiltonian 
\begin{align}
\lim\limits_{q\to 0} h_{dia} = \frac{e^2 {B^{z}}^2}{2m} x^2,	
\end{align}
from which (\ref{eq:metric_gab}) is derived, is proportional to the position operator squared, which is known to bound the spread of Wannier functions~\cite{Marzari1997,Souza2000}. The quantum metric also plays an important role in superfluids~\cite{Peotta2015,Torma2022} and in the capacitance of insulators~\cite{Komissarov2024,Verma2025}.  

Now, since we have set $\tau\to \infty$ in all denominators of (\ref{eq:n2g}), the $q\to 0$ limit raises the question of possible divergences, particularly from the first factor when $m=n$. In Appendixes \ref{sec:1q3}, \ref{sec:1q2}, and \ref{sec:1q1}, we show that this is not the case. In Appendix \ref{sec:n2_O1}, we calculate the response tensor $\varsigma^{(2)}$ as a sum: 
\begin{align}
\varsigma^{(2)} &= \varsigma_{11+21}^{(2)} + \bigg(\varsigma_{12+22}^{(2)} +\varsigma_{13+24}^{(2)} +\varsigma_{14+23}^{(2)}\bigg) + \varsigma_{15+25}^{(2)} \nn \\
&~~~+  \bigg(\varsigma_{31}^{(2)} + \varsigma_{32}^{(2)}\bigg).  
\label{eq:varsigma_sum}
\end{align}
The notation is explained in Appendix \ref{sec:1q3}. The splitting corresponds to different grouping of band indices: all three bands coinciding (first term), two (and only two) coinciding (next three terms), none coinciding (fifth term), and the diamagnetic contributions (last two terms).

The first term is the $m=n=l$ contribution and is analogous to the second term in the linear response (\ref{eqn:n1_berry_c}). Note that other one-band terms may arise from band resummations. As shown in Appendix \ref{sec:n2_O1}, the resulting expressions are quite  involved because they address the most general case of a multiband crystal with arbitrary symmetry. To build intuition, we now present specific examples.

\section{Example: Geometric response}
\label{sec:geometric}
There are terms in the nonlinear response which depend explicitly on the quantum metric. The origin of some of them is, intuitively,  the rotation of the Bloch wave function in the complex projective plane as the electron moves in k space. This mechanism is independent of spin or orbital angular momentum, and in this sense, it is a \textit{geometric} response. Here we identify two such contributions. First, we collect terms which depend explicitly on the quantum metric and on $f''_n$. These are easier to interpret since the factor multiplying $f''_n$ is necessarily a magnetic moment squared. These terms are the first term of \ref{eq:o1_11_21_v3}, seventh term of \ref{eq:o1_12_22_v2}, and last term of \ref{eq:o1_13_24_v1}, and their contribution for the arbitrary field direction is
\begin{align}
n^{(2)}_{geo} = \frac{1}{2V}\sum_{n\v{k}} f''_n B^c \left( \mathfrak{m}_{n}^{(2)cc'} + \bar{\mathfrak{m}}_{n}^{(2)cc'}\right) B^{c'},  
\end{align} 
where the quadratic \textit{geometric} moments are 
\begin{align}
\mathfrak{m}_{n}^{(2)cc'} &= e^2 \epsilon_{cab} \epsilon_{c'a'b'} g^{aa'}_{n} v^b_n v^{b'}_n, \\
\bar{\mathfrak{m}}_{n}^{(2)cc'} &= \frac{1}{2}\delta_{cc'} e^2 \epsilon_{cab} g^{aa}_{n} v^b_n v^{b}_n. 
\end{align} 
In Sec. \ref{sec:11_21_terms} we showed that in a simple model one of these dominates the intraband response.

\section{Example: magnetovolume effect}
\label{sec:magnetovolume}
From semiclassical electron dynamics, it is known that the k-space volume of a wave packet changes in the presence of a magnetic field~ \cite{Xiao2005}. A similar effect occurs in crystals. The particle number is fixed by the background ionic charge, but the crystal must maintain its periodic electronic density. Here, we assume an ideal infinite crystal with no boundaries; the case with boundaries will be analyzed elsewhere. In other words, the electronic density cannot accumulate in a particular region of the crystal. This means that the Brillouin zone volume must adjusts its size (perhaps its shape too) to comply with the change in average density. We can expand the volume in powers of the field as 
\begin{align}
V= V^{(0)} + v^{(1)a} B^a  + v^{(2)ab} B^a B^b + \cdots,  
\end{align}
and since $n=N/V$, the fractional change in volume is 
\begin{align}
\frac{V^{(1)}}{V^{(0)}} &= -\frac{n^{(1)}}{n^{(0)}} \label{eq:V1}, \\
\frac{V^{(2)}}{V^{(0)}} &= \left(\frac{V^{(1)}}{V^{(0)}}\right)^2 - \frac{n^{(2)}}{n^{(0)}}.
\label{eq:dv}
\end{align}
If TRS is present, $V^{(1)}=0$, but if TRS is broken, 
\begin{align}
\frac{n^{(0)}}{V^{(0)}} \left(\frac{\partial V}{\partial \v{B}}\right)_{\mu} = \frac{1}{e} \pmb{\sigma}_{H},
\end{align}
suggesting discontinuous jumps in the volume as a function of the magnetic field in the quantum Hall regime (assuming this expression remains valid in that limit). As the volume attempts to change, the elastic degrees of freedom resist, generating a pressure differential. The pressure can also be expanded in powers of the magnetic field as
\begin{align}
P= P^{(0)} + p^{(1)a} B^a  + p^{(2)ab} B^a B^b + \cdots,  
\end{align}
and assuming constant compressibility, 
\begin{align}
\kappa = \frac{1}{n}\frac{\partial n}{\partial P},
\end{align}
we obtain
\begin{align}
\kappa P^{(1)}  &= \frac{n^{(1)}}{n^{(0)}} \label{eq:P1}, \\
\kappa P^{(2)}  &= -\frac{1}{2} \left(\frac{n^{(1)}}{n^{(0)}}\right)^2 + \frac{n^{(2)}}{n^{(0)}}.
\label{eq:dP}
\end{align}
If TRS holds, $P^{(1)}=0$, but if TRS is broken,
\begin{align}
\kappa n^{(0)} \left(\frac{\partial P}{\partial \v{B}}\right)_{\mu} = -\frac{1}{e} \pmb{\sigma}_{H}. 
\end{align}
Magnetovolume effects such as these are well-known in invar alloys~\cite{Takahashi2013}, where they arise from itinerant spins and are usually studied using phenomenological models.


\section{Example: intraband contribution}
\label{sec:11_21_terms}
We explicitly calculate the first term in (\ref{eq:varsigma_sum}), given by (\ref{eq:o1_11_21_v3}), for an effective two-band model near the zone center: 
\begin{align}
h_{0\v{k}} = v\hbar (k^x \sigma^y - k^y \sigma^x). 
\label{eq:h0k}
\end{align}
We assume the Fermi energy lies in the conduction band. Had we started from (\ref{eq:h0k}) and applied a magnetic field, we would have opened an gap in the energy spectrum. Instead, we assume (\ref{eq:h0k}) is an effective model incorporating all such effects. For this simple model, the integrals can be evaluated analytically since the FS is a circle and $f''_n$ is localized near the FS. An explicit calculation shows that the largest contribution comes from the quantum metric tensor: 
\begin{align}
\varsigma^{(2)}_{11+21} &\approx -\frac{2}{V}\hspace{-2pt}\left(\frac{e}{2i}\right)^2 \sum_{n\v{k}} \frac{3}{2} f''_{n} g_{n}^{xx} v_n^y v_n^y, \\
&= \frac{e^2}{\hbar^2 k_F^2} \alpha, 
\label{eq:z2_intra_model}
\end{align}
where $\alpha=-9/256\pi=-.011$ is a dimensionless coefficient. The second-largest contribution in (\ref{eq:o1_11_21_v3}) corresponds to the 4th term, with a coefficient of $0.0012$--an order of magnitude smaller. The orbital magnetic moments vanish for this model, so the second and sixth terms are zero, as is the fifth term in (\ref{eq:o1_11_21_v3}). We can write this result in a more suggestive form as

\begin{align}
n^{(2)}_{11+21} & \approx n^{(0)} (\chi_F B^z)^2 4\pi^2 \alpha, 
\end{align}
where $\chi_F B^z = eB^z/\hbar k_F^2$ is a dimensionless measure of the geometric \textit{stiffness} of the FS. We use $n^{(0)} = k_F^2/4\pi^2$ as a density scale for this model. Considering a typical value of $k_F \sim 1$ nm$^{-1}$ and $B=1$ T, we obtain $\chi_F B = 1.5\times 10^{-3}$, and hence, 
\begin{align}
\frac{n^{(2)}_{11+21}}{n^{(0)}} &\sim\frac{1}{k_F^4} \sim -10^{-4} \%.
\label{eq:n2vn0}
\end{align}
In this example, the density decreases by a small percent due to the presence of the magnetic field. The fractional change scales as $\sim 1/k_F^4$, implying that smaller FSs are more sensitive magnetic fields. In general, the order of magnitude change in density is estimated from (\ref{eq:n2vn0}) but for a more precise value would require large-scale numerical simulations that include all terms in (\ref{eq:varsigma_sum}) and account for the detailed crystal structure of the material.

\section{Conclusion}
\label{sec:conclusion}
We have studied the average density of a clean 3D crystal in a static homogeneous magnetic field. We showed that for insulators, in   
the linear response, the average density agrees with the well-known Streda formula. For metals, or more generally, when particle occupations in k-space have discontinuities, there is an additional contribution arising from the orbital magnetic moments at the FS. This may have implications for itinerant ferromagnets in the anomalous quantum Hall regime (assuming the expression can be extrapolated to that case).

To quadratic order, we derived  a general expression for the second-order density response in terms of Bloch wave functions.   
 This expression is complex, as it accounts for the most general case of a multiband 3D crystal, with arbitrary symmetry, metallic or insulating. The expression includes FS, non-FS, one-band, two-band, and three-band contributions. The relative magnitude of these contributions should be analyzed on a case-by-case basis, ideally using first-principles numerical codes. Interestingly, the quantum metric tensor induces a magnetic moment due to rotations of the Bloch wave function in the complex projective plane. In a simple two-band model, we find that the quantum metric provides the dominant contribution to the intraband ($n=m=l$) response. We also showed  that density variations manifest corresponding change in volume or pressure and derived explicit expressions for these magnetovolume effects.  

The transverse resistivity as a function of magnetic field typically exhibits a linear response at small magnetic fields and a nonlinear behavior at larger fields ($\sim 1$ T). The carrier density is usually extracted from the slope in the linear regime using a semiclassical two-carrier model~\cite{Ziman2001}. The quadratic corrections to the density derived here may apply to the nonlinear regime of the transverse resistivity~\cite{comment1}. 

Finally, the method developed is free of singularities, explicitly gauge invariant, treats intraband and interband processes on equal footing, can accommodate relaxation, and can be extended to analyze electronic behavior well beyond semiclassical dynamics.

\section*{Acknowledgments}
I acknowledge support from NSF grant DMR-2015639.

\section*{Data Availability}
No data were created or analyzed in this study.

\appendix

\section{Density matrix}
\label{app:densitymat}
The single-particle Hamiltonian describing Bloch electrons in the presence of a magnetic field is 

\begin{align}
h&= \frac{1}{2m}(\v{p}-e \v{A})^2 + U \nn \\
 &= h_0-\frac{e}{2m}(\v{p}\cdot\v{A} +\v{A}\cdot\v{p}) + \frac{e^2 A^2}{2m},
\end{align}
where $h_0=p^2/2m + U$ is the unperturbed single-particle Hamiltonian whose eigenfunctions and eigenvalues are Bloch wave functions and band energies $\varphi_{n\v{k}}$ and $\varepsilon_{n\v{k}}$ (\ref{eq:bloch_eig_val_vec}). $U$ is the ionic periodic potential, $m$ is the mass of the electron, and $e<0$ is the electron's charge. The magnetic field is represented by the vector potential $\v{A}=\hat{\v{y}} (B^{z}/q) \sin (\v{g}\cdot \v{r})$, where the magnetic field is $\v{B}=\pmb{\nabla}\times \v{A} = \hat{\v{z}} B^{z} \cos(\v{g}\cdot \v{r})$ and $\v{g}=\hat{\v{x}} q$. We do not exclude the possibility of a time-dependent field. In terms of Bloch operators
\begin{align}
\hat{H}= \hat{H}_0 + \hat{W}^{(1)} + \hat{W}^{(2)},
\end{align}
where 
\begin{align}
\hat{H}_0 &= \sum_{n\v{k}} \varepsilon_{n\v{k}} a_{n\v{k}}^{\dagger} a_{n\v{k}}, \\
\hat{W}^{(1)} &= \sum_{n\v{k}m\v{k}'} W^{(1)}_{n\v{k}m\v{k}'}   a_{n\v{k}}^{\dagger} a_{m\v{k}'}, \\
\hat{W}^{(2)} &= \sum_{n\v{k}m\v{k}'} W^{(2)}_{n\v{k}m\v{k}'} a_{n\v{k}}^{\dagger} a_{m\v{k}'},
\end{align}
and $W^{(1)}$ and $W^{(2)}$ are the matrix elements of the first-- and second-order perturbations. The magnetic field is (possibly) time dependent:
\begin{align}
B^{z}= \sum_{\beta} B_{\beta}^{z} e^{-i(\omega_{\beta} + i0^{+})t},
\end{align}
where $\beta$ labels the frequency components of the field. We can factor this time dependence from the matrix elements as 
\begin{align}
W^{(1)}_{n\v{k}m\v{k}'} &\equiv \sum_{b\beta} \bar{W}^{(1)b\beta}_{n\v{k}m\v{k}'} B^{b}_{\beta} e^{-i\omega_{\beta} t}, \\
W^{(2)}_{n\v{k}m\v{k}'} &\equiv \sum_{b\beta c\sigma} \bar{W}^{(2)b\beta c\sigma}_{n\v{k}m\v{k}'} B^{b}_{\beta} B^{c}_{\sigma}  e^{-i\omega_{\Sigma} t}, 
\end{align}
where $\omega_{\Sigma} = \omega_{\beta} + \omega_{\sigma}$. Explicit calculation of the matrix elements gives 
\begin{align}
\bar{W}^{(1)z\beta}_{n\v{k}m\v{k}'} &= \frac{e}{4iq}(\delta_{\v{k},\v{k}'-\v{g}} - \delta_{\v{k},\v{k}'+\v{g}} ) 
\langle u_{n\v{k}} | \hat{v}^{y}_{\v{k}} + \hat{v}^{y}_{\v{k}'} |  u_{m\v{k}'}\rangle,  \\
\bar{W}^{(2)z\beta z \sigma}_{n\v{k}m\v{k}'} &= \frac{1}{2m}\bigg(\frac{e}{2iq}\bigg)^2
(\delta_{\v{k},\v{k}'+2\v{g}} \hspace{-2pt}-\hspace{-2pt} 2 \delta_{\v{k},\v{k}'} \hspace{-2pt}+\hspace{-2pt} \delta_{\v{k},\v{k}'- 2\v{g}}) \nn \\
&~~~~~~~~~~~~~~~~~~~~~~~~~~~~~~~\times  \langle u_{n\v{k}} |  u_{m\v{k}'}\rangle.  
\label{eq:W1W2}
\end{align}
The equation of motion of the density matrix in the Bloch basis,
\begin{align}
\rho_{m\v{k}'n\v{k}} \equiv \langle a_{n\v{k}}^{\dagger} a_{m\v{k}'}  \rangle,
\end{align}
is obtained in the usual way: 
\begin{align}
&\frac{\partial}{\partial t} \rho_{n'\v{k}'n\v{k}} + i\omega_{n'\v{k}'n\v{k}} \rho_{n'\v{k}'n\v{k}}  \nn \\
& ~~~~~~~~~~-\frac{1}{i\hbar} \sum_{m \v{p}} ( W_{n'\v{k}'m\v{p}} \rho_{m\v{p}n\v{k}} - \rho_{n'\v{k}'m\v{p}} W_{m\v{p}n\v{k}} ) \nn \\
& ~~~~~~~~~~=-\frac{1}{\tau}(\rho_{n'\v{k}'n\v{k}} - \rho^{(0)}_{n'\v{k}'n\v{k}}),
\label{eqn:eomrho}
\end{align} 
where $W \equiv W^{(1)} + W^{(2)}$.  We added a phenomenological collision integral in the relaxation time approximation on the right-hand side of (\ref{eqn:eomrho}). Now we look for solutions in powers of the magnetic field as  
\begin{align}
\rho = \rho^{(0)} + \rho^{(1)} + \rho^{(2)} \cdots,
\end{align}
subject to the initial condition
\begin{align}
\rho_{n'\v{k}'n\v{k}}(t=-\infty) =\rho_{n'\v{k}'n\v{k}}^{(0)} =  \delta_{nn'}\delta_{\v{k}\v{k}'} f_{n\v{k}}.
\end{align}

\subsection{Zero order}
To zero order, $\rho^{(0)}_{n'\v{k}'n\v{k}}=\delta_{nn'}\delta_{\v{k}\v{k}'} f_{n\v{k}}$  satisfies (\ref{eqn:eomrho}), where $f_{n\v{k}}$ is Fermi function (\ref{eq:Fermi_func}).

\subsection{First order}
To linear order (and long times) 
\begin{align}
\rho^{(1)}_{n'\v{k}'n\v{k}} = \sum_{\beta} \bar{\rho}^{(1)z\beta}_{n'\v{k}'n\v{k}}  B_{\beta}^{z} e^{-i\omega_{\beta}t} 
\label{eqn:rho1st_tot}
\end{align}
is a solution of (\ref{eqn:eomrho}) as long as 
\begin{align}
\bar{\rho}^{(1)b\beta}_{n'\v{k}'n\v{k}} &= \hbar^{-1}\bar{W}^{(1)b\beta}_{n'\v{k}'n\v{k}}\frac{f_{n'\v{k}'n\v{k}}}{\omega_{n'\v{k}'n\v{k}} - \bar{\omega}_{\beta}}.
\label{eqn:rho1st}
\end{align}
Here, we defined $\bar{\omega}_{\beta} \equiv \omega_{\beta} + i/\tau$. The density matrix satisfies $[\bar{\rho}^{(1)z\beta}_{n'\v{k}'n\v{k}}(\omega_{\beta})]^{*}= \bar{\rho}^{(1)z\beta}_{n\v{k}n'\v{k}'}(-\omega_{\beta})$, and hence, it is (as expected) Hermitian, $(\rho^{(1)}_{n'\v{k}'n\v{k}})^{*}= \rho^{(1)}_{n\v{k}n'\v{k}'}$.

\subsection{Second order}
To second order
\begin{align}
\rho^{(2)}_{n'\v{k}'n\v{k}} = \sum_{\beta\sigma} \bar{\rho}^{(2)b\beta c \sigma}_{n'\v{k}'n\v{k}}  B_{\beta}^{b} B_{\sigma}^{c} e^{-i\omega_{\Sigma}t}
\label{eqn:rho2}
\end{align}
is a solution of (\ref{eqn:eomrho}) as long as 
\begin{align}
\bar{\rho}^{(2)b\beta c \sigma}_{n'\v{k}'n\v{k}} &= \frac{\hbar^{-1}}{\omega_{n'\v{k}'n\v{k}} \hspace{-1pt}-\hspace{-1pt} \bar{\omega}_{\Sigma}} 
\sum_{m\v{p}} [\bar{\rho}^{(1)c \sigma}_{n'\v{k}'m\v{p}} \bar{W}^{(1)b\beta}_{m\v{p}n\v{k}} \nn \\
&~~~~~~~~~~~~~~~~~~~~~~~~~~~-\bar{W}^{(1)b\beta}_{n'\v{k}'m\v{p}} \bar{\rho}^{(1)c \sigma}_{m\v{p}n\v{k}}  ] \nn \\
&~~+\hbar^{-1}\bar{W}^{(2)b\beta c \sigma}_{n'\v{k}'n\v{k}}\frac{f_{n'\v{k}'n\v{k}}}{\omega_{n'\v{k}'n\v{k}} - \bar{\omega}_{\Sigma}}.
\label{eqn:bar_rho2}
\end{align}
We could proceed with finite frequency by symmetrizing with respect to exchange of indices $b\beta \leftrightarrow c\sigma$, etc. In this case a time-dependent electric field needs to be included to satisfy Faraday's law. Our modified Boltzmann equation method is readily suited to incorporate homogeneous electric and magnetic  fields, finite frequency, dissipation, and intraband and interband terms on equal footing. This ambitious program will be explored elsewhere. For now we set all frequencies to zero, and hence, the indices $\beta$ and $\sigma$ are absent; substituting (\ref{eqn:rho1st}) and setting $\tau \to \infty$ (clean limit), we obtain 
\begin{align}
\rho^{(2)}_{m\v{k}-\v{G}n\v{k}} &\hspace{-2pt}=\hspace{-2pt} \frac{- B^{b}B^{c}}{\varepsilon_{m\v{k}-\v{G}} \hspace{-2pt}-\hspace{-2pt}\varepsilon_{n\v{k}}} 
\sum_{l\v{p}} \left[\frac{f_{l\v{p}}-f_{n\v{k}}}{\varepsilon_{l\v{p}} -\varepsilon_{n\v{k}}} \hspace{-2pt}-\hspace{-2pt}
\frac{f_{m\v{k}-\v{G}} - f_{l\v{p}}}{\varepsilon_{l\v{k}-\v{G}} - \varepsilon_{l\v{p}}} \right] \nn \\
& ~~~~~~~~~~~~~~~~~~~~~~~~~~~~~~\times \bar{W}^{(1)b}_{m\v{k}-\v{G}l\v{p}} \bar{W}^{(1)c}_{l\v{p}n\v{k}}  \nn \\
&+ B^{b}B^{c} \bar{W}^{(2)b c}_{m\v{k}-\v{G}n\v{k}} \frac{f_{m\v{k}-\v{G}} \hspace{-2pt}-\hspace{-2pt} f_{n\v{k}}}{\varepsilon_{m\v{k}-\v{G}} \hspace{-2pt}-\hspace{-2pt}\varepsilon_{n\v{k}}}.
\label{eqn:rho2nd2}
\end{align}
Repeated indices are to be summed over.

\subsection{Symmetric Landau gauge}
We could use a more symmetric form of the vector potential $\v{A}=-\epsilon_{abc}\hat{\v{e}}_b (B^{a}/2q)\sin(q r^c)$. In this case, four Fourier modes contribute to the second-order response: $n_{2\v{g}^1},n_{2\v{g}^2},n_{\v{g}^1 + \v{g}^2},n_{\v{g}^1 - \v{g}^2}$, where $\v{g}^1$ and $\v{g}^2$ are an arbitrary pair of basis vectors in the reciprocal space perpendicular to a given magnetic field. For example, to linear response, the density matrix is the sum of the responses to each wave vector, 
\begin{align}
\bar{\rho}^{(1)j\beta}_{n'\v{k}'n\v{k}} &= \frac{1}{2}\epsilon_{jab} \frac{f_{n'\v{k}'n\v{k}}  \bar{W}^{j\beta}_{n'\v{k}'n\v{k}}}{\hbar(\omega_{n'\v{k}'n\v{k}} - \bar{\omega}_{\beta})}.
\label{eqn:rho1st_symm}
\end{align}
%

\section{$\varsigma^{(1)z}$}
\label{sec:eqn_n1_berry_c}
Separating the $m=n$ and $m\neq n$ terms and expanding to leading order in $q$, we obtain
\begin{align}
\varsigma^{(1)z} &=-\frac{e}{V} \textrm{Im}\sum_{n\v{k}}  \langle u_{n}| (\partial_{x} P_{n})\hat{v}^{y}_{\v{k}} | u_{n} \rangle f'_{n} \nn \\
&~~~- \frac{e}{V} \textrm{Im} \sum_{\begin{smallmatrix} nm\v{k} \\ n\neq m  \end{smallmatrix}} \langle u_{n}| (\partial_{x} P_{m})\hat{v}^{y}_{\v{k}} | u_{n} \rangle \frac{f_{m} - f_{n}}{\varepsilon_{m}-\varepsilon_{n}}.
\end{align}
Now we use (\ref{eq:qmt}) in the first term and (\ref{eqn:umk_vk_unk}) and  (\ref{eq:unk_complete}) in the second, and after some algebra, 
\begin{align}
\varsigma^{(1)z} &= \frac{1}{V} \sum_{n\v{k}} m_{n}^{z} f'_n -\frac{2e}{V\hbar} \textrm{Im}\sum_{n\v{k}}  \langle \partial_y u_{n}| \partial_x u_{n} \rangle f_{n}. 
\end{align}
Then using (\ref{eq:berry_curvature}) and (\ref{eqn:Hall_conductivity}) and deducing the result for arbitrary magnetic field direction, we get (\ref{eqn:n1_berry_c}).

\section{First \textit{sine} mode}
\label{app:sineterms}
In the series (\ref{eqn:n_f_exp}), the $sine$ mode, $n'_{\v{g}} sin(\v{g}\cdot \v{r})$, vanishes. This is reassuring since otherwise,  the density would depend on the form of $\v{A}$. To start we evaluate the matrix element
\begin{align}
n'_{\v{g}} &= \frac{2}{V}\int d\v{r}~ n \sin{( \v{g}\cdot\v{r})}
\label{eqn:F_modes_sine}
\end{align}
to obtain
\begin{align}
n'_{\v{g}} = -\frac{2}{V} \textrm{Im}\sum_{nm\v{k}} \langle u_{n\v{k}} | u_{m\v{k}-\v{g}} \rangle \rho_{m\v{k}-\v{g}n\v{k}}.
\label{eq:npg}
\end{align}
Now use (\ref{eqn:rho1st}) and separate interband from intraband terms and take the $q\to 0$ limit. Note that we only need to check $\sim 1/q$ terms in (\ref{eq:npg}) since higher-order terms vanish. Explicit calculation shows that to leading order
\begin{align}
n'_{\v{g}}= \frac{eB_{0}^{z}}{qV}\sum_{n\v{k}} \partial_{x} f_{n}=0,
\end{align}
and hence, there is no contribution from the first $sine$ mode in the Fourier series.

\section{$O(1/q^{3}$)}
\label{sec:1q3}
It is easy to check that an expansion in small $q$ of Fermi function differences over band energy differences (with same momenta), such as those inside the square brackets of (\ref{eq:n2g}), produces no extra $1/q$ or $1/(\partial_x \varepsilon_n)$ factors. This is true in general even when two or three bands are equal. However, when $m=n$, the first factor in the first term of (\ref{eq:n2g}) gives an overall $1/q$ factor and hence leads to an expression of $1/q^3$ order. $1/q^3$ diverges as $q \to 0$ unless (\ref{eq:n2g}) vanishes identically under general conditions. We now show that this is the case. 

To begin, let us separate the band sums in (\ref{eq:n2g}) into distinct groups: (1) all three bands $nml$ are equal, (2) two (and only two) bands are equal, and (3) no bands are equal; schematically,
\begin{align}
\sum_{nml} = \sum_{\textcolor{red}{nml}} + \left(\sum_{\textcolor{red}{nm}\textcolor{blue}{l}} 
+ \sum_{\textcolor{red}{n}\textcolor{blue}{m}\textcolor{red}{l}} + \sum_{\textcolor{blue}{n}\textcolor{red}{ml}}\right) + \sum_{\textcolor{red}{n}\textcolor{blue}{m}\textcolor{green}{l}}.
\label{eq:6sums}  
\end{align}
Matching colors indicates bands with equal indices. The second term in (\ref{eq:n2g}) has only two sums ($m=n$ and $m\neq n$): 
\begin{align}
\sum_{nm} = \sum_{\textcolor{red}{nm}} + \sum_{\textcolor{red}{n}\textcolor{blue}{m}}.
\label{eq:6sums_dia}  
\end{align}
Next, we label the first term inside the square brackets of (\ref{eq:n2g}) as  $\Circled{1}$, the second as $\Circled{2}$, and the diamagnetic term as $\Circled{3}$, in such a way that the $n=m=l$ term in $\Circled{1}$ is $\Circled{11}$, the term with $n=m$ in $\Circled{1}$ is $\Circled{12}$, and so on. Then Eq. (\ref{eq:n2g}) can be written schematically as 
\begin{align}
\varsigma_{2\v{g}}^{(2)zz} = \Circled{1} + \Circled{2} + \Circled{3},   
\end{align}
where
\begin{align}
\Circled{1} &= \Circled{11} + \Circled{12} + \Circled{13} +\Circled{14} + \Circled{15},   \\
\Circled{2} &= \Circled{21} + \Circled{22} + \Circled{23} +\Circled{24} + \Circled{25},   \\
\Circled{3} &= \Circled{31} + \Circled{32}.  
\end{align}
Finally, explicit calculation shows that to order $1/q^3$
\begin{align}
\left[\Circled{11} + \Circled{21}\right]_{1/q^3} &= 0, \\
\left[\Circled{12} + \Circled{22}\right]_{1/q^3} &= 0, \\
\Circled{rest}_{1/q^3}&=0. \label{eq:rest_invq3}
\end{align}
Equation (\ref{eq:rest_invq3}) means that each of the remaining terms of the form $\Circled{ab}_{1/q^3}$ vanish individually.

\section{$O(1/q^{2}$)}
\label{sec:1q2}
With the same notation as in Appendix \ref{sec:1q3}, explicit calculation of the $m=n=l$ term to order $1/q^2$ gives
\begin{align}
\left[\Circled{11} + \Circled{21}\right]_{1/q^2} &= \frac{1}{V}\hspace{-2pt}\left(\frac{e}{2iq}\right)^2\hspace{-1pt}\sum_{n\v{k}}  f''_n v^{y}_n v^{y}_n,  \\
&= -\frac{1}{V}\hspace{-2pt}\left(\frac{e}{2iq}\right)^2\hspace{-1pt}\sum_{n\v{k}} f'_n \left[\frac{1}{m} \hspace{-2pt}+\hspace{-2pt} \frac{2}{\hbar}\omega_{nl}r_{nl}^{y}r_{ln}^{y}\right].
\end{align}
To obtain the second line we used integration by parts and (\ref{eq:inv_mass_ten}). Similarly, the $m=n$ term is
\begin{align}
[\Circled{12} \hspace{-2pt}+\hspace{-2pt} \Circled{22}]_{1/q^2} &=\frac{2}{V}\hspace{-2pt}\left(\frac{e}{2iq}\right)^2\hspace{-1pt}\textrm{Re} \hspace{-2pt}\sum_{n\v{k}} \frac{-1}{(\partial_x \varepsilon_{n})2} \bigg[\frac{f_{nl}(\partial_x \varepsilon_{n})2}{(\varepsilon_{n}-\varepsilon_{l})^2} \nn \\
 &~~~~~~~~~~~~~~~~~~~~~~~~~~+ \frac{(-\partial_x f_n)2}{\varepsilon_{n}-\varepsilon_{l}} \bigg], \label{eq:12p22v1} \\ 
&= \frac{2}{V}\hspace{-2pt}\left(\frac{e}{2iq}\right)^2\hspace{-1pt}\textrm{Re}\sum_{n\v{k}} f'_n \frac{1}{\hbar}\omega_{nl}r_{nl}^{y}r_{ln}^{y} \label{eq:12p22v2}.
\end{align}
The first term in (\ref{eq:12p22v1}) vanishes by exchanging $n \leftrightarrow l$. Then, using (\ref{eqn:umk_vk_unk}) twice and (\ref{eq:pos_mat_ele}) on the remaining term, we get (\ref{eq:12p22v2}). Finally, the diamagnetic term is
\begin{align}
\Circled{31}_{1/q^2} &= \frac{1}{V}\hspace{-2pt}\left(\frac{e}{2iq}\right)^2\hspace{-1pt}\textrm{Re} \sum_{n\v{k}} f'_n \frac{1}{m}, 
\label{eq:31}
\end{align}
and one can check 
\begin{align}
\left[\Circled{11} + \Circled{21}\right]_{1/q^2}+ \left[\Circled{12} + \Circled{22}\right]_{1/q^2} + \Circled{31}_{1/q^2} &=0.
\end{align}
As expected, the diamagnetic term is necessary for particle conservation. Each of the remaining terms vanish independently,
\begin{align}
\Circled{rest}_{1/q^2}&=0.
\end{align}
%

\section{$O(1/q$)}
\label{sec:1q1}
To order $1/q$ the expansion of Eq.~(\ref{eq:n2g}) contains more terms. To begin, the $m=n=l$ contribution is
\begin{align}
[\Circled{11} +  &\Circled{21}]_{1/q} = \frac{2}{V}\hspace{-2pt}\left(\frac{e}{2iq}\right)^2\hspace{-2pt} q  \textrm{Re}\sum_{n\v{k}} \frac{1}{2} f''_n \nn \\
&~~~~~~~~~~~~~~~~\times \bigg [ -2 \langle u_n | (\partial_x P_n)\hat{v}^{y}_{\v{k}} P_n \hat{v}^{y}_{\v{k}}  | u_n \rangle  \nn \\
&~~~~~~~~~~~~~~~~~~~~~~~~ -\langle u_n |P_n (\partial_x \hat{v}^{y}_{\v{k}} P_n)\hat{v}^{y}_{\v{k}}  | u_n \rangle \bigg]  \nn \\
&~~~~~~~~~~~-\frac{2}{V}\hspace{-2pt}\left(\frac{e}{2iq}\right)^2\hspace{-2pt} q \textrm{Re}\sum_{n\v{k}} \frac{1}{2} f'''_n (\partial_x \varepsilon_n) v_n^y v_n^y \label{eq:11_21_invq1}  \\ 
&~~~~~~~~~=0. 
\label{eq:11_21_invq1_v2} 
\end{align}
Note that factors of the form $1/(\partial_x \varepsilon_{n})$, although ubiquitous, do not appear in (\ref{eq:11_21_invq1}). This will turn out to be the case for all expressions to $O(1)$. If we use (\ref{eq:partialvk}), the expression in the square brackets is $(-1)\partial_x[v_n^y v_n^y]$; then an integration by parts on the second term gives (\ref{eq:11_21_invq1_v2}). Similarly, the $m=n$ terms give
\begin{align}
[\Circled{12} +  &\Circled{22}]_{1/q} = \nn \\
&\frac{2}{V}\hspace{-2pt}\left(\frac{e}{2iq}\right)^2 q  \textrm{Re} \sum_{\begin{smallmatrix} nl\v{k} \\ n\neq l  \end{smallmatrix}}
 \frac{(-1)}{(\partial_x \varepsilon_{n})2} \nn \\
&~~~~~\times \left(\frac{f_{nl}(\partial_x \varepsilon_{n})2}{(\varepsilon_{n}-\varepsilon_{l})^2} +  \frac{(-\partial_x f_{n})2}{\varepsilon_{n}-\varepsilon_{l}} \right) \nn \\
&~~~~~\times \bigg[-\langle u_{n} | (\partial_x P_n) \hat{v}^{y}_{\v{k}} P_l \hat{v}^{y}_{\v{k}}| u_n \rangle 2 \nn \\
&~~~~~~~~~~~~~-\langle u_{n} | P_n (\partial_x \hat{v}^{y}_{\v{k}} P_l) \hat{v}^{y}_{\v{k}}| u_n \rangle  \bigg] \nn \\
&+\frac{2}{V}\hspace{-2pt}\left(\frac{e}{2iq}\right)^2 q  \textrm{Re} \sum_{\begin{smallmatrix} nl\v{k} \\ n\neq l  \end{smallmatrix}}  \frac{(-1)}{(\partial_x \varepsilon_{n})2} \nn \\
&~~~~~~ \times \bigg\{\frac{f_{nl}[4(\partial_x \varepsilon_{n})^2 - 4(\partial_x \varepsilon_{n})(\partial_x \varepsilon_{l})]}{(\varepsilon_{n}-\varepsilon_{l})^3} \nn \\
&~~~~~~~~~~~-\frac{f_{nl}2(\partial_x^2 \varepsilon_n)}{(\varepsilon_{n}-\varepsilon_{l})^2} \nn \\
&+\frac{4(-\partial_x f_n)(\partial_{x} \varepsilon_n) \hspace{-2pt}+\hspace{-2pt} 2(\partial_x f_n)(\partial_{x} \varepsilon_l) \hspace{-2pt}+\hspace{-2pt} 2(\partial_x f_l)(\partial_{x} \varepsilon_n)}{(\varepsilon_{n}-\varepsilon_{l})^2} \nn \\
&~~~~~~~~~~~+ \frac{2(\partial^2_x f_n)}{\varepsilon_{n}-\varepsilon_{l}} \bigg\} v_{nl}^y v_{ln}^y \nn \\
&+ \frac{2}{V}\hspace{-2pt}\left(\frac{e}{2iq}\right)^2 q  \textrm{Re} \sum_{\begin{smallmatrix} nl\v{k} \\ n\neq l  \end{smallmatrix}} \frac{(-1)(\partial_x^2 \varepsilon_n)}{2 (\partial_x \varepsilon_{n})^{2}} \nn \\
&~~~~~~\times \bigg[\frac{f_{nl}(\partial_x \varepsilon_{n})2}{(\varepsilon_{n}-\varepsilon_{l})^2} +\frac{2(-\partial_x f_n)}{\varepsilon_{n}-\varepsilon_{l}} \bigg]  v_{nl}^y v_{ln}^y \label{eq:qinv_12_22_v0} \\
&=-\frac{2}{V}\hspace{-2pt}\left(\frac{e}{2iq}\right)^2 q  \textrm{Re} \sum_{\begin{smallmatrix} nl\v{k} \\ n\neq l  \end{smallmatrix}} \partial_x\left(\frac{f'_n v_{nl}^y v_{ln}^y}{\varepsilon_n -\varepsilon_l} \right) \label{eq:qinv_12_22_v1}\\
&=0.
\end{align}
A few algebraic steps used to obtain (\ref{eq:qinv_12_22_v1}) are as follows: (1) The first summand inside the curly brackets of second term of (\ref{eq:qinv_12_22_v0}) is pure imaginary and hence vanishes when taking the real part; this can be seen by exchanging $n \leftrightarrow l$. (2) The second summand in the curly brackets of second term of (\ref{eq:qinv_12_22_v0}) adds to zero against the first summand inside square bracket of last term. (3) The second summand in the square brackets in the last term adds to zero against one term from the last summand in the curly brackets of second term. (4) The first summand inside the round brackets of first term vanishes as it is purely imaginary, and (5) we use (\ref{eq:partialvk}) in the second summand inside the round brackets of first term to simplify. Finally, collecting derivatives, we obtain (\ref{eq:qinv_12_22_v1}), which vanishes because the integrand is periodic in the Brillouin zone. Note that we assume the periodic gauge for Bloch wave functions, i.e., $\varphi_{n\v{k}+\v{G}} = \varphi_{n\v{k}}$ for $\v{G}$ in the reciprocal lattice.  Similar manipulations give
\begin{align}
\left[\Circled{14} + \Circled{23}\right]_{1/q} &= 0, \\
\left[\Circled{13} + \Circled{24}\right]_{1/q} &= 0, \\
\left[\Circled{15} + \Circled{25}\right]_{1/q} &= 0, \label{eq:15p25_qinv1} \\
\Circled{31}_{1/q}=\Circled{32}_{1/q}&= 0. 
\end{align}
In ($\ref{eq:15p25_qinv1}$) we used $F_{nml}=-F_{mnl}$, where 
\begin{align}
F_{nml}\equiv \frac{f_{ln}}{\varepsilon_{l}-\varepsilon_{n}} - \frac{f_{ml}}{\varepsilon_{m}-\varepsilon_{l}},
\label{eq:Fnml},
\end{align} 
to show that the sum is pure imaginary and hence vanishes upon taking the real part.

\section{$O(1)$}
\label{sec:n2_O1}
In this appendix we compute the response tensor $\varsigma^{(2)zz}$ by expanding (\ref{eq:n2g}) to leading order in small $q$. From Appendixes~\ref{sec:1q3}, \ref{sec:1q2}, and \ref{sec:1q1} we know the leading order is $O(1)$, as expected. Using the same notation as in Appendix \ref{sec:1q3}, we write 

\begin{align}
\varsigma^{(2)} &= \varsigma_{11+21}^{(2)} + \left(\varsigma_{12+22}^{(2)} +\varsigma_{13+24}^{(2)} +\varsigma_{14+23}^{(2)}\right) + \varsigma_{15+25}^{(2)} \nn \\
& +  \left(\varsigma_{31}^{(2)} + \varsigma_{32}^{(2)}\right). 
\end{align}
%

\subsection{$n=m=l$ terms}
We first calculate the $n=m=l$ term. After some algebra we obtain
\begin{align}
 \varsigma_{11+21}^{(2)zz} &= -\frac{2}{V}\hspace{-2pt}\left(\frac{e}{2i}\right)^2   \textrm{Re} \sum_{n\v{k}} (-1)\frac{1}{2} f''_{n} \nn \\
&~~~~~~~~~~~~~~\times \bigg[\hspace{9pt}2 \langle u_n| (\partial_x^2 P_n) \hat{v}_{\v{k}}^y P_n \hat{v}_{\v{k}}^y | u_n \rangle \nn \\
&~~~~~~~~~~~~~~~~~~ + 2 \langle u_n| (\partial_x P_n) (\partial_x \hat{v}_{\v{k}}^y P_n) \hat{v}_{\v{k}}^y | u_n \rangle \nn \\
&~~~~~~~~~~~~~~~~~~ + \frac{1}{2} \langle u_n|P_n (\partial^2_x \hat{v}_{\v{k}}^y P_n) \hat{v}_{\v{k}}^y | u_n \rangle \bigg] \nn\\
&\hspace{11pt}-\frac{2}{V}\hspace{-2pt}\left(\frac{e}{2i}\right)^2    \sum_{n\v{k}} \frac{5}{12} f''_{n}\partial_x[(\partial_x v^y_n)v^y_n ] \nn \\
&\hspace{11pt}-\frac{2}{V}\hspace{-2pt}\left(\frac{e}{2i}\right)^2    \sum_{n\v{k}} \frac{1}{8} f''_{n}\partial_y[(\partial_x v^x_n)v^y_n ]. \label{eq:o1_11_21_v1}
\end{align}
Few comments are in order: (1) (\ref{eq:o1_11_21_v1}) is explicitly gauge invariant. (2) All terms depend on $f''_n = \partial^2 f_n/\partial \varepsilon_n^2$, which is highly localized near the FS (if there is one). (3) From dimensional analysis the integrands in (\ref{eq:o1_11_21_v1}) have the form $\sim f''_n \times$ \textit{magnetic moment}$^2$, where the moments are formed from derivatives of $(P_n \hat{v}^y)(P_n \hat{v}^y)$ and $v_n^a v_n^b$, and (4) factors of the form $1/(\partial_x \varepsilon_n)$ do not appear in the final expression, and hence, (\ref{eq:o1_11_21_v1}) is free of divergences arising from flat (or nearly flat) bands. 

We can write the expression in square brackets in the first term of (\ref{eq:o1_11_21_v1}) in a more familiar form. Using (\ref{eq:projector}), (\ref{eq:partialvk}), (\ref{eqn:partial_a_vbnk}) and its second derivative [(\ref{eqn:vk_on_unk})], (\ref{eq:qgeoten2}), and the QMT (\ref{eq:qmt}), we obtain
\begin{align}
 \varsigma_{11+21}^{(2)zz} &= -\frac{2}{V}\hspace{-2pt}\left(\frac{e}{2i}\right)^2 \sum_{n\v{k}} \frac{1}{2} f''_{n} \nn \\
&~~~~~~~~\times \hspace{-2pt}\bigg[~~~3 g_{n}^{xx} v_n^y v_n^y  +  2 \xi_{n}^x v_n^y \frac{1}{e} m_{n}^z \nn \\
&~~~~~~~~~~~~- (\partial_x^2 v_n^y) v_n^y - \frac{3}{4} (\partial_x v_{n}^y)^2 \nn\\ 
&~~~~~~~~~~~~- \frac{1}{\hbar} \textrm{Re} \langle \partial_x^2 u_n| \varepsilon_n - h_{0\v{k}}| \partial_y u_{n}\rangle v_n^y \nn \\
&~~~~~~~~~~~~+\frac{1}{e^2} (m_n^{z})^2 \bigg] \nn\\
&\hspace{11pt}-\frac{2}{V}\hspace{-2pt}\left(\frac{e}{2i}\right)^2 \sum_{n\v{k}} \frac{5}{12} f''_{n}\partial_x[(\partial_x v^y_n)v^y_n ] \nn \\
&\hspace{11pt}-\frac{2}{V}\hspace{-2pt}\left(\frac{e}{2i}\right)^2 \sum_{n\v{k}} \frac{1}{8} f''_{n}\partial_y[(\partial_x v^x_n)v^y_n ]. \label{eq:o1_11_21_v2}
\end{align}
If we combine the last two terms, 
\begin{align}
 \varsigma_{11+21}^{(2)zz} &= -\frac{2}{V}\hspace{-2pt}\left(\frac{e}{2i}\right)^2 \sum_{n\v{k}} \frac{1}{2} f''_{n} \nn \\
&~~~~~~~~\times \hspace{-2pt}\bigg[\hspace{10pt} 3 g_{n}^{xx} v_n^y v_n^y  +  2 \xi_{n}^x v_n^y \frac{1}{e} m_{n}^z \nn \\
&~~~~~~~~~~~ +\frac{1}{24} \partial_x^2(v_n^y v_n^y) + \frac{1}{4} (\partial_x v_{n}^x)(\partial_y v_{n}^y) \nn \\ 
&~~~~~~~~~~~ -\frac{1}{\hbar} \textrm{Re} \langle \partial_x^2 u_n| \varepsilon_n - h_{0\v{k}}| \partial_y u_{n}\rangle v_n^y \nn \\
&~~~~~~~~~~~ +\frac{1}{e^2} (m_n^{z})^2 \bigg]. 
\label{eq:o1_11_21_v3}
\end{align}
On the right-hand side we recognize distinct forms of angular momentum squared made from basic objects such as velocity, metric tensor $g_{n}^{xx}$, position $\xi^x_{n}$ (\ref{eq:pos_mat_ele_intra}), and orbital moment $m_n^z$ and less familiar forms such as the fifth 
 term. In particular, the contribution form the last term, 
\begin{align}
 n^{(2)}_{11+21,6th} &= \frac{1}{4V}\sum_{n\v{k}} f''_{n} (\v{m}_n\cdot\v{B}) (\v{m}_n\cdot \v{B}), 
\label{eq:o1_11_21_v4}
\end{align}
is the obvious extension of $(1/V)\sum_{n\v{k}}f_n'\v{m}\cdot \v{B}$ in (\ref{eqn:n1_berry_c}) to the nonlinear regime, and the  contribution from the first term, 
\begin{align}
 n^{(2)}_{11+21,1st} &= \frac{3 e^2}{4V}\sum_{n\v{k}} f''_{n}  g^{xx}_n v_n^y v_n^y B^z B^z, 
\label{eq:o1_11_21_v5}
\end{align}
is a geometric response induced by the rotation of the Bloch wave functions in the complex projective plane. It is independent of spin or orbital angular momentum. We see similar contributions in seventh term of \ref{eq:o1_12_22_v2} and last term of \ref{eq:o1_13_24_v1}.

\subsection{$n=m$ terms}
Next, we compute the $n=m$ term in (\ref{eq:6sums}), by far the most involved. The expressions can be greatly simplified to show that factors of the form $1/(\partial_x \varepsilon_n)$ cancel. This is a good point to stop simplifying since expressions lacking explicit singularities are amenable to simulation in large-scale numerical codes. The expression is  
\begin{align}
\varsigma_{12+22}^{(2)zz} &= \nn\\
&-\frac{2}{V}\hspace{-2pt}\left(\frac{e}{2i}\right)^2 \textrm{Re} \sum_{\begin{smallmatrix} nl\v{k} \\ n\neq l  \end{smallmatrix}} \bigg[\frac{f_{nl}}{(\varepsilon_{n}\hspace{-2pt}-\hspace{-2pt}\varepsilon_{l})^2} \hspace{-2pt}-\hspace{-2pt} \frac{f'_n}{\varepsilon_{n}\hspace{-2pt}-\hspace{-2pt}\varepsilon_{l}} \bigg] \nn \\
&~~~~~~~~~~~~~~\times \bigg[\hspace{9pt} 2 \langle u_n| (\partial_x^2 P_n) \hat{v}_{\v{k}}^y P_l \hat{v}_{\v{k}}^y | u_n \rangle \nn \\
&~~~~~~~~~~~~~~~~~~ +  2\langle u_n| (\partial_x P_n) (\partial_x \hat{v}_{\v{k}}^y P_l) \hat{v}_{\v{k}}^y | u_n \rangle \nn \\
&~~~~~~~~~~~~~~~~~~ + \frac{1}{2} \langle u_n|P_n (\partial_x^2 \hat{v}_{\v{k}}^y P_l) \hat{v}_{\v{k}}^y | u_n \rangle \bigg] \nn\\
&-\frac{2}{V}\hspace{-2pt}\left(\frac{e}{2i}\right)^2 \textrm{Re} \sum_{\begin{smallmatrix} nl\v{k} \\ n\neq l  \end{smallmatrix}} 
\frac{f'_n}{\varepsilon_{n}\hspace{-2pt}-\hspace{-2pt}\varepsilon_{l}} \partial_x^2 [v_{nl}^y v_{ln}^y] \nn \\
&+\frac{2}{V}\hspace{-2pt}\left(\frac{e}{2i}\right)^2 \textrm{Re} \sum_{\begin{smallmatrix} nl\v{k} \\ n\neq l  \end{smallmatrix}} 
(\partial_x f'_n) \partial_x \bigg(\frac{1}{\varepsilon_{n}\hspace{-2pt}-\hspace{-2pt}\varepsilon_{l}}\bigg) v_{nl}^y v_{ln}^y \nn \\
&-\frac{2}{V}\hspace{-2pt}\left(\frac{e}{2i}\right)^2 \textrm{Re} \sum_{\begin{smallmatrix} nl\v{k} \\ n\neq l  \end{smallmatrix}} 
\frac{f_{nl}}{(\varepsilon_{n}\hspace{-2pt}-\hspace{-2pt}\varepsilon_{l})^3} \nn \\
&~~~~~~~~\times \bigg[\frac{(\partial_x \varepsilon_{n})(\partial_x \varepsilon_{n} - 6\partial_x \varepsilon_{l})}{(\varepsilon_{n}\hspace{-2pt}-\hspace{-2pt}\varepsilon_{l})^4} \hspace{-2pt}-\hspace{-2pt} \partial_x^2 \varepsilon_n \bigg] v_{nl}^y v_{ln}^y \nn \\
&+\frac{2}{V}\hspace{-2pt}\left(\frac{e}{2i}\right)^2 \textrm{Re} \sum_{\begin{smallmatrix} nl\v{k} \\ n\neq l  \end{smallmatrix}} 
\frac{f'_n}{(\varepsilon_{n}\hspace{-2pt}-\hspace{-2pt}\varepsilon_{l})^2} \nn\\
&~~~~~~~~~~~~~~~~~~~~~~~~\times \bigg[\frac{(\partial_x \varepsilon_{l})^2}{\varepsilon_{n}\hspace{-2pt}-\hspace{-2pt}\varepsilon_{l}} \hspace{-2pt}+\hspace{-2pt} \frac{1}{2}\partial_x^2 \varepsilon_l \bigg] v_{nl}^y v_{ln}^y \nn \\
&-\frac{2}{V}\hspace{-2pt}\left(\frac{e}{2i}\right)^2 \textrm{Re} \sum_{\begin{smallmatrix} nl\v{k} \\ n\neq l  \end{smallmatrix}} 
(\partial_x f_n) \partial_x \bigg[\frac{1}{(\varepsilon_{n}\hspace{-2pt}-\hspace{-2pt}\varepsilon_{l})^2}\bigg] v_{nl}^y v_{ln}^y \nn \\
&-\frac{2}{V}\hspace{-2pt}\left(\frac{e}{2i}\right)^2 \textrm{Re} \sum_{\begin{smallmatrix} nl\v{k} \\ n\neq l  \end{smallmatrix}} 
\frac{1}{2}(\partial_x^2 f_n) \frac{1}{(\varepsilon_{n}\hspace{-2pt}-\hspace{-2pt}\varepsilon_{l})^2} v_{nl}^y v_{ln}^y \nn \\
&-\frac{2}{V}\hspace{-2pt}\left(\frac{e}{2i}\right)^2 \textrm{Re} \sum_{\begin{smallmatrix} nl\v{k} \\ n\neq l  \end{smallmatrix}} 
\frac{f'_n}{(\varepsilon_{n}\hspace{-2pt}-\hspace{-2pt}\varepsilon_{l})^2} (\partial_x^2 \varepsilon_{n}) v_{nl}^y v_{ln}^y \nn \\
&+\frac{2}{V}\hspace{-2pt}\left(\frac{e}{2i}\right)^2 \textrm{Re} \sum_{\begin{smallmatrix} nl\v{k} \\ n\neq l  \end{smallmatrix}} 
\frac{2}{3} \frac{f'''_n}{\varepsilon_{n}\hspace{-2pt}-\hspace{-2pt}\varepsilon_{l}} (\partial_x \varepsilon_n)^2 v_{nl}^y v_{ln}^y \nn \\
&+\frac{2}{V}\hspace{-2pt}\left(\frac{e}{2i}\right)^2 \textrm{Re} \sum_{\begin{smallmatrix} nl\v{k} \\ n\neq l  \end{smallmatrix}} 
\frac{f''_n}{\varepsilon_{n}\hspace{-2pt}-\hspace{-2pt}\varepsilon_{l}} (\partial_x^2 \varepsilon_{n}) v_{nl}^y v_{ln}^y 
\label{eq:o1_12_22_v1}
\end{align}
To obtain (\ref{eq:o1_12_22_v1}) we observed that various terms are pure imaginary, different parts of expressions cancel each other (or combine), and integration by parts must be used at some points. Note that (1) again, (\ref{eq:o1_12_22_v1}) is explicitly gauge invariant, (2) the ubiquitous factors $1/(\partial_x \varepsilon_n)$ arising from the expansion of $1/(\varepsilon_{n\v{k}\hspace{-1pt}-\hspace{-1pt}2\v{g}} \hspace{-2pt}-\hspace{-2pt}\varepsilon_{n\v{k}})$ in (\ref{eq:n2g}) do not appear explicitly, and (3) all but two terms are explicitly FS terms. 

The last four terms can be converted to one-band terms by band resummation. For example, if we use (\ref{eqn:umk_vk_unk}), (\ref{eq:unk_complete}), (\ref{eqn:vk_on_unk}), and (\ref{eq:qmt}), the last term can be written as
\begin{align}
\varsigma_{12+22,last}^{(2)zz} &= \frac{2}{V}\hspace{-2pt}\left(\frac{e}{2i}\right)^2  \sum_{ n\v{k} } 
f''_n (\partial_x v^{x}_n)\frac{1}{2} \left[\partial_y v^{y}_n - \frac{\hbar}{m}\right].
\end{align}
Similar manipulations give
\begin{align}
\varsigma_{12+22}^{(2)zz} &= \nn\\
&-\frac{2}{V}\hspace{-2pt}\left(\frac{e}{2i}\right)^2 \textrm{Re} \sum_{\begin{smallmatrix} nl\v{k} \\ n\neq l  \end{smallmatrix}} \bigg[\frac{f_{nl}}{(\varepsilon_{n}\hspace{-2pt}-\hspace{-2pt}\varepsilon_{l})^2} \hspace{-2pt}-\hspace{-2pt} \frac{f'_n}{\varepsilon_{n}\hspace{-2pt}-\hspace{-2pt}\varepsilon_{l}} \bigg] \nn \\
&~~~~~~~~~~~~~~\times \bigg[\hspace{9pt} 2 \langle u_n| (\partial_x^2 P_n) \hat{v}_{\v{k}}^y P_l \hat{v}_{\v{k}}^y | u_n \rangle \nn \\
&~~~~~~~~~~~~~~~~~~ +  2\langle u_n| (\partial_x P_n) (\partial_x \hat{v}_{\v{k}}^y P_l) \hat{v}_{\v{k}}^y | u_n \rangle \nn \\
&~~~~~~~~~~~~~~~~~~ + \frac{1}{2} \langle u_n|P_n (\partial_x^2 \hat{v}_{\v{k}}^y P_l) \hat{v}_{\v{k}}^y | u_n \rangle \bigg] \nn\\
&-\frac{2}{V}\hspace{-2pt}\left(\frac{e}{2i}\right)^2 \textrm{Re} \sum_{\begin{smallmatrix} nl\v{k} \\ n\neq l  \end{smallmatrix}} 
\frac{f'_n}{\varepsilon_{n}\hspace{-2pt}-\hspace{-2pt}\varepsilon_{l}} \partial_x^2 [v_{nl}^y v_{ln}^y] \nn \\
&+\frac{2}{V}\hspace{-2pt}\left(\frac{e}{2i}\right)^2 \textrm{Re} \sum_{\begin{smallmatrix} nl\v{k} \\ n\neq l  \end{smallmatrix}} 
(\partial_x f'_n) \partial_x \bigg(\frac{1}{\varepsilon_{n}\hspace{-2pt}-\hspace{-2pt}\varepsilon_{l}}\bigg) v_{nl}^y v_{ln}^y \nn \\
&-\frac{2}{V}\hspace{-2pt}\left(\frac{e}{2i}\right)^2 \textrm{Re} \sum_{\begin{smallmatrix} nl\v{k} \\ n\neq l  \end{smallmatrix}} 
\frac{f_{nl}}{(\varepsilon_{n}\hspace{-2pt}-\hspace{-2pt}\varepsilon_{l})^3} \nn \\
&~~~~~~~~\times \bigg[\frac{(\partial_x \varepsilon_{n})(\partial_x \varepsilon_{n} - 6\partial_x \varepsilon_{l})}{(\varepsilon_{n}\hspace{-2pt}-\hspace{-2pt}\varepsilon_{l})^4} \hspace{-2pt}-\hspace{-2pt} \partial_x^2 \varepsilon_n \bigg] v_{nl}^y v_{ln}^y \nn \\
&+\frac{2}{V}\hspace{-2pt}\left(\frac{e}{2i}\right)^2 \textrm{Re} \sum_{\begin{smallmatrix} nl\v{k} \\ n\neq l  \end{smallmatrix}} 
\frac{f'_n}{(\varepsilon_{n}\hspace{-2pt}-\hspace{-2pt}\varepsilon_{l})^2} \nn\\
&~~~~~~~~~~~~~~~~~~~~~~~~\times \bigg[\frac{(\partial_x \varepsilon_{l})^2}{\varepsilon_{n}\hspace{-2pt}-\hspace{-2pt}\varepsilon_{l}} \hspace{-2pt}+\hspace{-2pt} \frac{1}{2}\partial_x^2 \varepsilon_l \bigg] v_{nl}^y v_{ln}^y \nn \\
&-\frac{2}{V}\hspace{-2pt}\left(\frac{e}{2i}\right)^2 \textrm{Re} \sum_{\begin{smallmatrix} nl\v{k} \\ n\neq l  \end{smallmatrix}} 
(\partial_x f_n) \partial_x \bigg[\frac{1}{(\varepsilon_{n}\hspace{-2pt}-\hspace{-2pt}\varepsilon_{l})^2}\bigg] v_{nl}^y v_{ln}^y \nn \\
&-\frac{2}{V}\hspace{-2pt}\left(\frac{e}{2i}\right)^2  \sum_{n \v{k}} 
\frac{1}{2}(\partial_x^2 f_n) \frac{1}{\hbar^2} g^{yy}_n \nn \\
&-\frac{2}{V}\hspace{-2pt}\left(\frac{e}{2i}\right)^2  \sum_{n\v{k}} 
f'_n (\partial_x^2 \varepsilon_n) \frac{1}{\hbar^2} g^{yy}_n\nn \\
&+\frac{2}{V}\hspace{-2pt}\left(\frac{e}{2i}\right)^2  \sum_{n\v{k}} 
\frac{1}{3} (\partial_x f''_n ) v^x_n \left[\partial_y v^{y}_n - \frac{\hbar}{m}\right] \nn \\
&+\frac{2}{V}\hspace{-2pt}\left(\frac{e}{2i}\right)^2  \sum_{ n\v{k} } 
f''_n (\partial_x v^{x}_n)\frac{1}{2} \left[\partial_y v^{y}_n - \frac{\hbar}{m}\right]
\label{eq:o1_12_22_v2}
\end{align}
Finally, we note the first term in (\ref{eq:o1_12_22_v2}) represents simply interband matrix elements involving $(P\hat{v}^y)(P\hat{v}^y)$.

\subsection{$n=l$ terms}
Similar calculations for $n=l$ term in (\ref{eq:6sums}) give
\begin{align}
\varsigma_{13+24}^{(2)zz} &= -\frac{2}{V}\hspace{-2pt}\left(\frac{e}{2i}\right)^2   \textrm{Re} \sum_{\begin{smallmatrix} nm\v{k} \\ n\neq m  \end{smallmatrix}} \frac{2f'_n}{\varepsilon_{m}\hspace{-2pt}-\hspace{-2pt}\varepsilon_{n}} \nn \\
&~~~~~~~~~~~~~~\times \bigg[\hspace{6pt} \langle u_n| (\partial_x^2 P_m) \hat{v}_{\v{k}}^y P_n \hat{v}_{\v{k}}^y | u_n \rangle \nn \\
&~~~~~~~~~~~~~~~~~ +  \langle u_n| (\partial_x P_m) (\partial_x \hat{v}_{\v{k}}^y P_n) \hat{v}_{\v{k}}^y | u_n \rangle \bigg] \nn\\
&+\frac{2}{V}\hspace{-2pt}\left(\frac{e}{2i}\right)^2   \textrm{Re} \sum_{\begin{smallmatrix} nm\v{k} \\ n\neq m  \end{smallmatrix}} \frac{2f'_m}{\varepsilon_{m}-\varepsilon_{n}} \nn \\
&~~~~~~~~~~~~~~\times \bigg[\hspace{6pt} \langle u_n| (\partial_x^2 P_m) \hat{v}_{\v{k}}^y P_m \hat{v}_{\v{k}}^y | u_n \rangle \nn \\
&~~~~~~~~~~~~~~~~~ +  \langle u_n| (\partial_x P_m) (\partial_x \hat{v}_{\v{k}}^y P_m) \hat{v}_{\v{k}}^y | u_n \rangle \bigg] \nn\\
&-\frac{2}{V}\hspace{-2pt}\left(\frac{e}{2i}\right)^2  \textrm{Re}  \sum_{\begin{smallmatrix} nm\v{k} \\ n\neq m  \end{smallmatrix}} 4 \frac{1}{\varepsilon_m\hspace{-2pt}-\hspace{-2pt}\varepsilon_n} \frac{(\partial_y f_{mn})}{\varepsilon_{m}\hspace{-2pt}-\hspace{-2pt}\varepsilon_{n}} \nn \\ 
&~~~~~~~~~~~~~~~~~~~~~~~~~~~~~\times\langle u_n |\partial_x u_n \rangle v_{mn}^y v_m^x \nn \\
&+\frac{2}{V}\hspace{-2pt}\left(\frac{e}{2i}\right)^2    \sum_{n\v{k}} 2 f''_{n} g_n^{xy} v^y_n v^x_n. 
\label{eq:o1_13_24_v1}
\end{align}
A few comments are in order: (1) (\ref{eq:o1_13_24_v1}) is explicitly gauge invariant, (2) the integrands are localized at the FS (if there  is one), and (3) the last term is a one-band term (after band resummation) proportional to the quantum metric.

\subsection{$m=l$ terms}
Next, the $m=l$ term in (\ref{eq:6sums}) is 
\begin{align}
\varsigma_{14+23}^{(2)zz} &= -\frac{2}{V}\hspace{-2pt}\left(\frac{e}{2i}\right)^2   \textrm{Re} \sum_{\begin{smallmatrix} nm\v{k} \\ n\neq m  \end{smallmatrix}} \frac{2f_{mn}}{(\varepsilon_{m}-\varepsilon_{n})^2} \nn \\
&~~~~~~~~~~~~~~\times \bigg[\hspace{5pt} \langle u_n| (\partial_x^2 P_m) \hat{v}_{\v{k}}^y P_m \hat{v}_{\v{k}}^y | u_n \rangle \nn \\
&~~~~~~~~~~~~~~~~~ +  \langle u_n| (\partial_x P_m) (\partial_x \hat{v}_{\v{k}}^y P_m) \hat{v}_{\v{k}}^y | u_n \rangle \bigg] \nn\\
&\hspace{11pt}+\frac{2}{V}\hspace{-2pt}\left(\frac{e}{2i}\right)^2   \textrm{Re} \sum_{\begin{smallmatrix} nm\v{k} \\ n\neq m  \end{smallmatrix}} \frac{2f_{mn}}{(\varepsilon_{m}-\varepsilon_{n})^2} \nn \\
&~~~~~~~~~~~~~~\times \bigg[\hspace{5pt} \langle u_n| (\partial_x^2 P_m) \hat{v}_{\v{k}}^y P_n \hat{v}_{\v{k}}^y | u_n \rangle \nn \\
&~~~~~~~~~~~~~~~~~ +  \langle u_n| (\partial_x P_m) (\partial_x \hat{v}_{\v{k}}^y P_n) \hat{v}_{\v{k}}^y | u_n \rangle \bigg] \nn\\
&\hspace{11pt}-\frac{2}{V}\hspace{-2pt}\left(\frac{e}{2i}\right)^2  \textrm{Re}  \sum_{\begin{smallmatrix} nm\v{k} \\ n\neq m  \end{smallmatrix}} 4 \partial_y \bigg(\frac{1}{\varepsilon_m\hspace{-2pt}-\hspace{-2pt}\varepsilon_n} \bigg) \frac{f_{mn}}{\varepsilon_{m}\hspace{-2pt}-\hspace{-2pt}\varepsilon_{n}} \nn \\ 
&~~~~~~~~~~~~~~~~~~~~~~~~~~\times\langle u_n |\partial_x u_n \rangle v_{mn}^y v_m^x \nn \\
&\hspace{11pt}+\frac{2}{V}\hspace{-2pt}\left(\frac{e}{2i}\right)^2  \textrm{Re}  \sum_{\begin{smallmatrix} nm\v{k} \\ n\neq m  \end{smallmatrix}} 4\bigg[\frac{f_{nm}}{(\varepsilon_m\hspace{-2pt}-\hspace{-2pt}\varepsilon_n)^2}\hspace{-2pt}-\hspace{-2pt} \frac{f'_n}{\varepsilon_m\hspace{-2pt}-\hspace{-2pt}\varepsilon_n} \bigg] \nn \\
&~~~~~~~~~~~~~~~~~~~~\times \langle u_n |\partial_x u_n \rangle (\partial_x \varepsilon_n) v_{mn}^y v_m^y. 
\label{eq:o1_14_23_v1}
\end{align}
The first and second terms in (\ref{eq:o1_14_23_v1}) are almost the same except for a band index difference.

\subsection{$n\neq m\neq l$ terms}
Next, the purely interband terms in (\ref{eq:6sums}) are 
\begin{align}
\varsigma_{15+25}^{(2)zz} &=-\frac{2}{V}\hspace{-2pt}\left(\frac{e}{2i}\right)^2   \textrm{Re} \sum_{\begin{smallmatrix} nml\v{k} \\ n\neq m\neq l \end{smallmatrix}} \frac{2F_{mnl}}{\varepsilon_{m}-\varepsilon_{n}} \nn \\
&~~~~~~~~~~~~~~~~~~~~\times\bigg[\hspace{5pt}\langle u_n| (\partial_x^2 P_m) \hat{v}_{\v{k}}^y P_l \hat{v}_{\v{k}}^y | u_n \rangle \nn \\
&~~~~~~~~~~~~~~~~~~~~~~~ +\langle u_n| (\partial_x P_m) (\partial_x \hat{v}_{\v{k}}^y P_l) \hat{v}_{\v{k}}^y | u_n \rangle \bigg] \nn\\
&\hspace{11pt}+\frac{2}{V}\hspace{-2pt}\left(\frac{e}{2i}\right)^2 \hspace{-2pt}\textrm{Re} \hspace{-8pt} \sum_{\begin{smallmatrix} nm\v{k} \\ n\neq m\neq l  \end{smallmatrix}} \hspace{-7pt}\frac{(-4)}{\varepsilon_m\hspace{-2pt}-\hspace{-2pt}\varepsilon_n} \bigg[\frac{f_{nl}}{(\varepsilon_m\hspace{-2pt}-\hspace{-2pt}\varepsilon_l)^2}\hspace{-2pt}-\hspace{-2pt} \frac{f'_n}{\varepsilon_m\hspace{-2pt}-\hspace{-2pt}\varepsilon_l} \bigg] \nn \\
&~~~~~~~~~~~~~~~~~~~~~~~~~\times (\partial_x \varepsilon_n)\langle u_n | \partial_x u_m \rangle v^{y}_{ml}v^{y}_{ln} \nn \\
&\hspace{11pt}-\frac{2}{V}\hspace{-2pt}\left(\frac{e}{2i}\right)^2  \textrm{Re} \hspace{-5pt}  \sum_{\begin{smallmatrix} nml\v{k} \\ n\neq m \neq l \end{smallmatrix}}  (-4) \frac{F_{nml}}{(\varepsilon_{m}-\varepsilon_{n})^2} \nn \\
&~~~~~~~~~~~~~~~~~~~~\times (\partial_x \varepsilon_m)\langle u_n | \partial_x u_m \rangle v^{y}_{ml}v^{y}_{ln}
\label{eq:o1_15_25_v1}
\end{align}
where $F_{nml}$ is given in (\ref{eq:Fnml}).

\subsection{Diamagnetic terms}
Finally, the diamagnetic intraband and interband terms in (\ref{eq:6sums}) are
\begin{align}
\varsigma_{31}^{(2)zz} &= -\frac{1}{V}\hspace{-2pt}\left(\frac{e}{2i}\right)^2   \sum_{n\v{k}}\frac{1}{m}\bigg[4 f'_n g^{xx}_n \hspace{-2pt}-\hspace{-2pt} \frac{1}{3} f''_n (\partial_x^2 \varepsilon_n) \bigg],  
\label{eq:o1_31_v1}  \\
\varsigma_{32}^{(2)zz} &= \frac{1}{V}\hspace{-2pt}\left(\frac{e}{2i}\right)^2  \sum_{nm\v{k}} \frac{1}{m} \frac{4f_{mn}}{\varepsilon_m \hspace{-2pt}-\hspace{-2pt}\varepsilon_n} r_{nm}^x r_{mn}^x.
\label{eq:o1_32_v1}
\end{align}
To obtain (\ref{eq:o1_31_v1}) we used two derivatives of (\ref{eq:projector}), two derivatives of (\ref{eq:un_norm}), and (\ref{eq:qgeoten}), and an integration by parts. To obtain (\ref{eq:o1_32_v1}) we used two derivatives of (\ref{eq:projector}) and (\ref{eq:pos_mat_ele}). Note the position matrix element squared in (\ref{eq:o1_32_v1}). This is expected since this term is nothing but the expectation value of the position operator squared in a Bloch state. Less obvious is (\ref{eq:o1_31_v1}), which is also the (intraband) expectation value of the position operator squared. One key feature of our method is that it recovers, on equal footing, both intraband and interband processes.

It is possible to use (\ref{eq:inv_mass_ten}) to eliminate the mass in favor of velocity and position matrix elements and combine the result with (\ref{eq:o1_11_21_v3}).

\section{Notation and identities}
\label{app:identities}
To avoid cumbersome notation, we often omit the crystal momentum index $\v{k}$ in expressions which are diagonal in $\v{k}$. Summation over a repeated variable index is usually assumed. We hope these and other notation details are clear from the context.
\begin{align}
h_{0} &= \frac{\v{p}^2}{2m} + U, \\
h_{0} \varphi_{n\v{k}}    &= \varepsilon_{n\v{k}} \varphi_{n\v{k}} \label{eq:bloch_eig_val_vec}, \\
h_{0\v{k}}          &\equiv e^{-i\v{k}\cdot\v{r}} h_{0} e^{i\v{k}\cdot\v{r}},  \\
h_{0\v{k}} |u_{n\v{k}}\rangle &= \varepsilon_{n\v{k}} |u_{n\v{k}}\rangle, \label{eqn:Hk_on_unk}\\
\varphi_{n\v{k}}          &= u_{n\v{k}} e^{i\v{k}\cdot\v{r}}, \\
|\varphi_{n\v{k}}\rangle  &\equiv |n\v{k} \rangle =  |u_{n\v{k}}\rangle e^{i\v{k}\cdot\v{r}}, \\
P_{n\v{k}}                &\equiv  |u_{n\v{k}}\rangle \langle u_{n\v{k}} |, \label{eq:projector}  \\
1                         &= \sum_{n} P_{n\v{k}}, \label{eq:unk_complete} 
\end{align}

\begin{align}
\epsilon_{abc} &=\textit{Levi-Civita symbol in 3D}, \\
\partial_a &\equiv \frac{\partial}{\partial k^a}, \label{eq:partial_def}\\
\partial_a^n &\equiv \frac{\partial}{\partial k^a}\cdots \frac{\partial}{\partial k^a} ~~~~(n~\textrm{times}),\label{eq:partial_def2}\\
\nabla^{a} &\equiv \frac{\partial}{\partial r^a}, \label{eq:nabla_def} 
\end{align}

\begin{align}
\delta_{n,l} &= \langle u_{n\v{k}}| u_{l\v{k}}\rangle, \label{eq:un_norm} \\
0 &= \langle u_{n\v{k}}| \partial_a u_{l\v{k}}\rangle + \langle \partial_a u_{n\v{k}}|u_{l\v{k}}\rangle, \label{eqn:id1} \\
r^{a}_{nm} &\equiv i\langle u_{n\v{k}} |\partial_a u_{m\v{k}} \rangle ~~~~~~ n\neq m \nn \\
& \equiv 0 ~~~~~~~~~~~~~~~~~~~~~~ n= m, \label{eq:pos_mat_ele} \\
\xi^{a}_{n\v{k}} &\equiv i\langle u_{n\v{k}} |\partial_a u_{n\v{k}} \rangle, \label{eq:pos_mat_ele_intra} \\
f_{n\v{k}} &\equiv f(\varepsilon_{n\v{k}}) \nn \\
& = [e^{(\varepsilon_{n\v{k}}-\mu)/k_B T}+1]^{-1} ~~\textrm{as}~~ T\to 0, \label{eq:Fermi_func}\\
f'_{n\v{k}} &\equiv \frac{\partial f_{n\v{k}}}{\partial \varepsilon_{n\v{k}}}, \\
\hbar\omega_{n\v{k}}&\equiv \varepsilon_{n\v{k}}, \\
\omega_{n\v{k}l\v{k}'} &\equiv \omega_{n\v{k}} - \omega_{l\v{k}'}, \\
f_{n\v{k}l\v{k}'}  &\equiv f_{n\v{k}}- f_{l\v{k}'},
\end{align}

\begin{align}
\hat{v}^{a}   &= -\frac{1}{m}i\hbar \nabla^a, \\
\hat{v}^{a}_{\v{k}} &\equiv e^{-i\v{k}\cdot\v{r}} \hat{v}^{a} e^{i\v{k}\cdot\v{r}} = \frac{1}{\hbar}\partial_a h_{0\v{k}}, 
\label{eq:vk}\\
\partial^{n}_{b} \hat{v}^{a}_{\v{k}} &= 0 ~~~~~~a\neq b, ~n\in \mathbb{N}, \label{eq:partialvk} \\
v_{nl}^{a}  &= \langle n\v{k}| \hat{v}^{a} |l\v{k} \rangle =  \langle u_{n\v{k}}| \hat{v}^a_{\v{k}} |u_{l\v{k}} \rangle,  \\
v^{a}_{n\v{k}} &\equiv v_{nn}^{a}(\v{k}) = \frac{1}{\hbar}\partial_{a}\varepsilon_{n\v{k}}, \label{eqn:vnk} \\
\partial_b v^{a}_{n\v{k}} &= \langle \partial_b u_{n}| \hat{v}^{a}_{\v{k}} | u_{n} \rangle + \langle u_{n}| \hat{v}^{a}_{\v{k}} | \partial_b u_{n} \rangle, ~~~a\neq b,	\label{eqn:partial_a_vbnk} 
\end{align}

\begin{align}
\hat{v}^{a}_{\v{k}} |u_{n\v{k}} \rangle &= \frac{1}{\hbar} (\varepsilon_{n\v{k}} \hspace{-2pt}-\hspace{-2pt} h_{0\v{k}} ) 
| \partial_a u_{n\v{k}}\rangle  +  v^{a}_{n\v{k}} |u_{n\v{k}} \rangle,  \label{eqn:vk_on_unk}\\
\langle u_{l\v{k}} |\hat{v}^a_{\v{k}} |u_{n\v{k}} \rangle &= \frac{1}{\hbar} (\varepsilon_{n\v{k}} \hspace{-2pt}-\hspace{-2pt} \varepsilon_{l\v{k}} ) \langle u_{l\v{k}}| \partial_{a} u_{n\v{k}}\rangle + v^{a}_{n\v{k}} \delta_{ln}, \label{eqn:umk_vk_unk}  \\
( \partial_i \hat{v}^{j}_{\v{k}}  ) |u_{n\v{k}}\rangle  &= 
(v^{j}_{n\v{k}} \hspace{-2pt}-\hspace{-2pt} \hat{v}^{j}_{\v{k}}) |  \partial_i u_{n\v{k}}  \rangle  
+ (i \leftrightarrow j) \nn \\ 
\hspace{-2pt}+\hspace{-2pt} &\frac{1}{\hbar} (\varepsilon_{n\v{k}} \hspace{-2pt}-\hspace{-2pt} h_{0\v{k}}) |  \partial_{ij} u_{n\v{k}}  \rangle 
\hspace{-2pt}+\hspace{-2pt} ( \partial_{i} v^{j}_{n\v{k}} ) |u_{n\v{k}}\rangle, \label{eqn:dvk_on_unk} \\
\langle u_{n\v{k}} |  ( \partial_{i} \hat{v}^{j}_{\v{k}} ) |u_{n\v{k}} \rangle &= 
-\frac{1}{\hbar} \langle \partial_{i} u_{n\v{k}}  |\varepsilon_{n\v{k}} \hspace{-2pt}-\hspace{-2pt} h_{0\v{k}}
| \partial_{j} u_{n\v{k}}  \rangle \hspace{-2pt}\nn \\
&~~+ (i \leftrightarrow j) +  \partial_{i} v^{j}_{n\v{k}},  \label{eqn:unk_dvk_unk} 
\end{align}

\begin{align}
\frac{1}{\hbar}\partial_{ab}\varepsilon_{n} &= \partial_{a} v^{b}_{n} =\partial_{b} v^{a}_{n} \nn \\	
&=\frac{\hbar}{m}\delta_{a,b} + \sum_{l}\omega_{nl} (r_{nl}^{a}r_{ln}^{b} + r_{nl}^{b}r_{ln}^{a}), \label{eq:inv_mass_ten} \\
\Omega^{ab}_{n\v{k}} &= -2\textrm{Im}\langle \partial_{a} u_{n\v{k}} | \partial_{b} u_{n\v{k}} \rangle \label{eqn:berry_potential} \\
&=i \sum_{l} (r_{nl}^a r^{b}_{lm} -r_{nl}^a r^{b}_{lm}) \\
\Omega^{a}_{n\v{k}} &=\frac{1}{2}\sum_{bc}\epsilon_{abc} \Omega^{bc}_{n\v{k}} \\
& = i\epsilon_{abc} \langle \partial_{b} u_{n\v{k}} | \partial_{c} u_{n\v{k}}, \rangle, \label{eq:berry_curvature} \\
\pmb{\sigma}_{H} &\equiv  \frac{e^2}{\hbar V} \sum_{n\v{k}} f_{n\v{k}} \pmb{\Omega}_{n\v{k}},  \label{eqn:Hall_conductivity} \\
m^a_{n\v{k}} &= \frac{ie}{2\hbar}\epsilon_{abc} \langle \partial_{b} u_{n\v{k}} | \varepsilon_{n\v{k}} \hspace{-2pt}-\hspace{-2pt} h_{0\v{k}}| \partial_{c} u_{n\v{k}} \rangle, \label{eqn:orb_moment} \\
\mathcal{M}^{a}_{n\v{k}} &= \frac{i e}{2\hbar} \epsilon_{abc} \langle \partial_{b} u_{n\v{k}}|\varepsilon_{n\v{k}}\hspace{-2pt}+\hspace{-2pt} h_{0\v{k}}| \partial_{c} u_{n\v{k}} \rangle, \label{eqn:orb_moment2} \\
Q^{ab}_{n\v{k}} &= \langle \partial_{a} u_{n\v{k}}| 1-P_{n\v{k}} | \partial_{b} u_{n\v{k}} \rangle,  \label{eq:qgeoten} \\
\textrm{Re} ~Q^{ab}_{n\v{k}} &= g^{ab}_{n\v{k}}, \label{eq:qgeoten2} \\
\textrm{Im} ~Q^{ab}_{n\v{k}} &= -\frac{1}{2}\Omega_{n\v{k}}^{ab}, \label{eq:qgeoten3}
\end{align}
Most of these definitions and identities are well-known or can be easily derived. Here, we comment only on few of them: (1) Eq.~(\ref{eqn:vk_on_unk}) follows from a $\v{k}$ derivative of (\ref{eqn:Hk_on_unk}), (\ref{eq:vk}), and (\ref{eqn:vnk}), (2) Eq.~(\ref{eqn:umk_vk_unk}) follows from (\ref{eqn:vk_on_unk}), (3) Eq.~(\ref{eqn:dvk_on_unk}) follows from taking two $\v{k}$ derivatives of (\ref{eqn:Hk_on_unk}), and (4) Eq.~(\ref{eqn:unk_dvk_unk}) follows from (\ref{eqn:dvk_on_unk}).


%

\end{document}